\documentclass[11pt]{article}
\pdfoutput=1

\usepackage{color}
\usepackage{amsmath, amsfonts,amssymb,theorem,euscript,
array,enumerate,amsfonts,mathrsfs}
\usepackage[T1]{fontenc}
\usepackage[latin1]{inputenc}
\usepackage{hhline}
\usepackage{graphicx}
\usepackage{eurosym}

\usepackage{dsfont}

\theoremheaderfont{\bfseries} \theoremstyle{plain}
\theorembodyfont{\upshape}

\newcommand{\nc}{\newcommand}
\nc{\ind}{\mathds{1}}

\newcommand{\R}{\mathbb{R}}

\newcommand{\E}{\mathcal{E}}
\newcommand{\F}{\mathcal{F}}

\DeclareMathOperator{\esssup}{esssup}

\def\esssup_#1{\underset{#1}{\mathrm{ess\,sup\, }}}
\def\essinf_#1{\underset{#1}{\mathrm{ess\,inf\, }}}
\def\argmax_#1{\underset{#1}{\mathrm{arg\,max\, }}}
\def\argmin_#1{\underset{#1}{\mathrm{arg\,min\, }}}

\def \ep{\hbox{ }\hfill$\Box$}

\def \Frac{\displaystyle\frac}

\def\b1{\bf 1}

\def\Dt#1{\Frac{\partial #1}{\partial t}}

\def\Dx#1{\Frac{\partial #1}{\partial x}}
\def\Dd#1{\Frac{\partial #1}{\partial d}}
\def\Dy#1{\Frac{\partial #1}{\partial y}}

\def\Dy#1{\Frac{\partial #1}{\partial y}}

\def\Dyy#1{\Frac{\partial^2 #1}{\partial y^2}}

\def\Dyd#1{\Frac{\partial^2 #1}{\partial y \partial d}}
\def\Ddd#1{\Frac{\partial^2 #1}{\partial d^2}}

\def \R{\mathbb{R}}

\def \E{\mathbb{E}}
\def \F{\mathbb{F}}

\def \P{\mathbb{P}}

\def \Ac{{\cal A}}

\def \Ec{{\cal E}}
\def \Fc{{\cal F}}

\def \ep{\hbox{ }\hfill$\Box$}

\newtheorem{Theorem}{Theorem}[part]

\newtheorem{Proposition}{Proposition}[part]

\newtheorem{Remark}{Remark}[part]

\def\reff#1{{\rm(\ref{#1})}}

\def\beqs{\begin{eqnarray*}}
\def\enqs{\end{eqnarray*}}
\def\beq{\begin{eqnarray}}
\def\enq{\end{eqnarray}}

\begin{document}

\title{An optimal trading  problem  in intraday electricity markets
\footnote{This study was supported by FiME (Finance for Energy Market Research Centre) and
the ``Finance et D\'eveloppement Durable - Approches Quantitatives'' EDF - CACIB Chair.  
The authors would like to thank Marc Ringeisen, Head of EDF R\&D Osiris Department for insightful discussion on trading and intraday market. 
}}
\author{
Ren\'e Aïd
\footnote{EDF R\&D and Finance for Energy Market Research Centre www.fime-lab.org, \sf rene.aid at edf.fr}
\qquad
Pierre Gruet
\footnote{LPMA, Universit\'e Paris Diderot,  \sf gruet at math.univ-paris-diderot.fr}
\qquad
Huy\^en Pham
\footnote{LPMA, Universit\'e Paris-Diderot and CREST-ENSAE, \sf  pham at math.univ-paris-diderot.fr}
}

\maketitle

\begin{abstract}
We consider  the problem of  optimal trading for a power producer  in the context of intraday electricity markets. 
The aim is to minimize the imbalance cost induced by the random residual demand in electricity, i.e. the consumption from the clients minus the 
production from renewable energy.  For a simple linear price 
impact model and a quadratic criterion, we explicitly obtain approximate  optimal strategies in the intraday market and thermal power generation, and 
exhibit some remarkable properties of the trading rate.  Furthermore, we study  the case  when there are jumps on the demand forecast and 
on the intraday price, typically due to error in the prediction of wind power generation. Finally, we  solve 
the problem when taking into account  delay constraints in thermal power  production.
\end{abstract}

\vspace{5mm}


\vspace{5mm}


\vspace{5mm}


\newpage

\section{Introduction}

The development of renewable energy sources in Europe as a response to global climate change has led to an increase of exchange in the intraday electricity markets. For instance, the exchanged volume on the European Energy Exchange (EEX) for Germany has grown from 2 TWh in 2008 to 25 TWh in 2013. This increase is mainly due to the level of forecasting error of wind production,  which leads  power producers owning  a large share of wind production to turn more than ever to intraday markets in order to adjust their position and avoid penalties for their imbalances. The accuracy of forecasts for  renewable power production from wind and solar may vary considerably depending on the agreggation level (local vs regional forecast) and the time horizon. For a complete survey on this problem, the reader can consult Giebel et al. \cite{Giebel11}, and may  have in mind that the root mean square error (RMSE) of the error forecast for the production of a wind farm in six hours can reach 20\% of its installed capacity. Many different intraday markets have been designed and are subject to different sets of regulation. But, in all cases, intraday markets offer power producer the possibility to buy or sell power for the next (say) 
9 hours to 32 hours (case of the French electricity market of EpexSpot). These trades can occur after the closing of the day-ahead market or during the clearing phase of the day-ahead market.  

The problem of   trading management  in  the intraday electricity  market for a balancing purpose has already drawn the attention in the literature. Henriot \cite{Henriot14} studied the problem of how the intraday market can help a power producer to deal with the wind production error forecast in a stylized discrete time model. In his model, the power producer is a wind producer who is trying to minimize her  sourcing cost  on the intraday market while maintaining a balance position between her  forecast production and her sales. Henriot's model takes into account the impact of the wind power producer on the intraday price with a deterministic inverse demand function, and  the intraday price is not a risk factor. The only risk factor comes from the error forecast of the wind production and its auto-correlation. Garnier and Madlener  \cite{Garnier14} studies the trade-off between entering into a deal in the intraday market right now and postponing it in a discrete time decision model where intraday prices follow a  geometric Brownian model and wind production error forecast follows an arithmetic Brownian motion. In their framework, the power producer is supposed to have no impact on intraday prices. Liquidity risk is taken into account as a probability of not finding a counter-party at the next trading window.

In this paper, we consider a power producer having at disposal some renewable energy sources (e.g. wind and solar), and thermal plants (e.g. coal, gas, oil, and nuclear sources), and who can buy/sell energy in the intraday markets. Her purpose is to minimize the  
imbalance cost, i.e. the cost induced by the difference between the demand of her clients minus the electricity produced and traded,  plus the production and trading costs.  In contrast with thermal power plants whose generation can be controlled, the power generated from renewable sources 
is subject to non controllable fluctuations or risks (wind speed, weather forecast) 
and is then considered here as a random factor just like the demand.  
We then call {\em the residual demand} the  demand minus the energy generated by renewable energy. 
Thus, the problem of the power producer is to minimize the imbalance costs arising from her residual demand by relying both on her own controllable thermal assets and on the intraday market. 
As in \cite{Garnier14}, we assume that the power producer has access to a continuously updated forecast of the residual demand to be satisfied at terminal date $T$ and that this forecast evolves randomly.  Moreover, the intraday price for delivery at time $T$ evolves also randomly and is correlated with the residual demand forecast. However, compared with  \cite{Garnier14}, the intraday market can be used for optimization purposes. We develop a model that allows us to study how power producers can take advantage of the interaction between the dynamics of the residual demand forecast and the dynamics of the intraday prices. 

Our model shares some links  with  optimal order execution problems, as  introduced in the seminal paper by Almgren and Chriss \cite{almcri00}, and then largely studied 
in the recent literature, see e.g. the survey paper \cite{schsly11}. 
In our context, the original feature with respect to this literature is the consideration of a random demand target and the possibility for the agent to use her thermal power production.  
This connection with optimal execution  is fruitful in the sense that it allows us to take into account several features of intraday markets while maintaining the tractability of the model sufficiently high to allow analytical solutions.  Hence, we take into account liquidity risk through a market impact, both permanent and temporary, on the electricity  price generated by a  power producer when 
trading in the intraday market.   
As in optimal execution problems, this impact is always in the adverse direction: when the producer sells, the price  decreases and when she buys, the price increases.  
Our setting is a continuous-time decision problem representing the possibility for the producer to make a deal at each time she wants and not only at pre-specified windows. Moreover,  it is  general enough  as it permits  us  to study the limiting cases of a pure retailer (no production function), a pure trader (no demand commitment) and an integrated player (player owning both clients and generation), small or large.

The main goal of this paper  is to derive analytical results, which provide explicit solutions  for the  (approximate)  optimal control,  hence giving enlightening economic interpretations of the optimal trading strategies.  In order to achieve such  analytical tractability, we have to make some simplifying assumptions on  the dynamics of the price process and of the residual demand forecast, as well as on the cost function, assumed to be of quadratic form  meaning  a simple linear growth of the marginal cost of production with respect to the production level.  We first consider a simple model for  a continuous price process with linear impact,  and demand forecast driven by an arithmetic Brownian motion, and neglect in a first step the delay of production when using thermal power plants. We then study an auxiliary control problem by relaxing the 
nonnegativity constraint  on the generation level,  for which we are able to derive explicit solutions.  The approximation error induced by this relaxation constraint is analyzed.  In next steps, we consider more realistic situations and investigate two extensions: (i) On one hand, we incorporate the case where the residual demand forecast is subject to sudden changes, related to prediction error for wind or solar power production, which may be quite 
important due to the difficulties for estimating wind speed and forecasting weather,  see \cite{benetal14}. This is formalized by jumps in the dynamics of the demand process, and consequently also on the price process.  Again, we are able to obtain  explicit solutions. Actually, the key tool in the derivation of all these analytical results is a suitable treatment  of the linear-quadratic structure of our stochastic control problem.  (ii) On the other hand, we introduce  natural delay constraints  in the production, and show how the optimal decision problem can be explicitly solved by a suitable reduction to a problem without delay.

 Our (approximate) optimal trading strategies present some remarkable properties. When the intraday price process is  a martingale, the optimal trading rate inherits the martingale property, which implies in particular that the net position of electricity shares has a  constant growth rate on average.  Moreover,  the optimal strategy consists in making at each time the forecast marginal cost  equal to the forecast intraday price. 
 This property follows the common sense of intraday traders. Consequently,  if the producer has made sales or purchases on the day-ahead such that her forecast marginal cost equals the day-ahead price and if the initial condition of the intraday price is the day-ahead price, thus, on average, the producer optimal trading rate is zero. This fact is no longer true when the demand forecast and the price follow processes with jumps. In this case, the optimal trading rate is a supermartingale or a submartingale depending on the relative probability and size of positive and negative jumps on the price process.  
 For this reason, contrary to  the case without jumps, the power producer may need to have a non-zero initial trading rate even if she has made sales or purchases on the day-ahead such that her forecast marginal cost equals the day-ahead price and if the initial condition of the intraday price is the day-ahead price.   We also quantify explicitly the impact of delay in production on the trading strategies. When the price process is a martingale, the net inventory in electricity shares  grows linearly on average, with 
 a change of slope (which is smaller) at  the time decision for the  production.

The outline of the paper is organized as follows.  We formulate the optimal trading pro\-blem in Section 2.  In Section 3, we study the optimal trading problem without delay.  We first solve explicitly the auxiliary optimal execution problem, and then study  the approximation on the solution to the original problem, by  focusing in particular on the error asymptotics.  We illustrate our results with some numerical tests and simulations.  We extend in Section 4  our results to the case where jumps in demand forecast  may arise. In Section 5, we show how the optimal trading problem with delay in production can be reduced to 
a problem without delay, and then leads to explicit solutions. Finally, the appendix  collects the explicit derivations of our solutions, which are justified by verification theorems.

\section{Problem formulation}

\setcounter{equation}{0}
\setcounter{Assumption}{0} \setcounter{Theorem}{0}
\setcounter{Proposition}{0} \setcounter{Corollary}{0}
\setcounter{Lemma}{0} \setcounter{Definition}{0}
\setcounter{Remark}{0}

We consider an agent on an intraday energy market, who is required to guarantee her  equilibrium supply/demand for a given fixed time $T$: she has to satisfy the demand of her customers by purchase/sale of energy on the intraday market at time $T$ and also by means of her thermal power generation.  We denote by $X_t$  the net position of sales/purchases of electricity at time $t$ $\leq$ $T$ for a delivery at terminal time $T$, assumed to be described by an absolutely continuous trajectory up to time $T$, and  by $q_t$ $=$ $\dot X_t$ the trading rate:  $q_t$ $>$ $0$ means an instantaneous purchase of electricity, while $q_t$ $<$ $0$ represents an instantaneous sale at time~$t$: 
\beq \label{dynX}
X_t &=& X_0 + \int_0^t q_s ds, \;\;\; 0 \leq t \leq T. 
\enq
Given the trading rate, the transactions occur with a market price impact: 
\beqs
P_t(q) &=& \hat P_t + \int_0^t g(q_s) ds + f(q_t). 
\enqs
Here, $(\hat P_t)_t$ is the unaffected intraday electricity price process on a filtered proba\-bility space $(\Omega,\Fc,\F=(\Fc_t)_{t\in [0,T]},\P)$,  
carrying some part of randomness of the market, and following the terminology in the seminal paper by Almgren and Chriss \cite{almcri00}, the term $f(q_t)$  refers to the temporary price impact, while $\int_0^t g(q_s)ds$ describes the permanent price impact. The price $(\hat P_t)_t$ may be seen as a forward price, evolving in real time, for delivery at time $T$. Let us then denote by $Y$ the intraday electricity price 
impacted by the past trading rate  $q$ of the agent, defined by:
\beqs
Y_t &:= & \hat P_t + \int_0^t g(q_s) ds.  
\enqs
We assume that  $Y_t$ is observable and quoted, which means actually that  the agent is a large trader and electricity producer, whose actions directly impact the intraday electricity price. The case where the agent is a small producer can be also dealt with by simply considering a 
zero permanent impact function $g$ $\equiv$ $0$. 
Notice that the transacted price is equal to the sum of the quoted price $Y$ and the temporary price impact:
\beq \label{relPY}
P_t(q) &=& Y_t + f(q_t).  
\enq
The residual demand $D_T$ is the consumption of clients of the agent minus the production from renewable energy at terminal date $T$, 
and we assume that the agent has access to 
a continuously updated forecast $(D_t)_t$  of the residual demand.  The agent can use her thermal power production  with a quantity $\xi$ at cost $c(\xi)$ in order to match as close as possible the target demand $D_T$. In practice, generation of electricity can not be obtained instantaneously and needs a delay to reach a required level of production. Hence, the decision to produce a quantity $\xi$ should be taken at time $T-h$, where $h$ $\in$ $[0,T]$ is the delay.  Thus, for  a controlled trading rate $q$ $=$ $(q_t)_t$ $\in$ $\Ac$, the set of real-valued  $\F$-adapted processes satisfying some integrability conditions to be precised later,  a production quantity $\xi$ $\in$ $L^0_+(\Fc_{T-h})$,  the set of nonnegative $\Fc_{T-h}$-measurable random variables, the total cost is: 
\beq 
 \int_0^T q_t P_t(q) dt + C(D_T-X_T,\xi) 
& := &  \int_0^T q_t P_t(q) dt + c(\xi) + \frac{\eta}{2} (D_T-X_T - \xi)^2.  \label{defC}
\enq
The first term in \reff{defC} represents the total running cost arising from the trading in the intraday electricity market, and the last term, where $\eta$ $>$ $0$, represents the quadratic penalization when the net position in sales/purchases of electricity 
$X_T + \xi$ (including the production quantity  $\xi$ at cost $c(\xi)$) at terminal date $T$ does not fit the effective demand $D_T$.  
 The objective of the agent is then to minimize over $q$ and $\xi$ the expected total cost: 
 \beq \label{defoptim}
 \mbox{ minimize over } q \in \Ac, \; \xi \in L^0_+(\Fc_{T-h}) & & \E \Big[  \int_0^T q_t P_t(q) dt  + C(D_T-X_T,\xi) \Big]. 
 \enq
 
 \begin{Remark} \label{rem1}
 {\rm {\bf 1)} The penalization term in the objective function above is a simplification of the effective penalization process that can be found in real electricity markets. For instance, the penalization of imbalances in the French electricity market depends both on the sign of the imbalance of the electricity system and on the price of imbalances (see \cite[chap 2., sec. 2.2.1]{Aid15}). Nevertheless, the positive coefficient $\eta$ captures the main objective of the penalization process. The agent has no incentive of being either too long or too short.
 
\vspace{1mm} 

\noindent {\bf 2)} On real markets, trading ends some time before the date of delivery, at which the agent has to ensure equilibrium (e.g. on the French electricity market, there is a delay of 45 minutes). We do not include that practical fact in our framework, by considering that the delay is null for the sake of clarity. There is no mathematical consequence: it is enough to have in mind that the delivery and production do not really take place at $T$, but at $T$ plus some delay.

\vspace{1mm} 
 
\noindent  {\bf 3)} The larger is $\eta$, the stronger is the incentive for the agent  to be as close as possible to the equilibrium supply-demand. At the limit, when  $\eta$ goes to infinity, the agent is formally constrained to fit supply and demand. However, the limiting problem when $\eta$ $=$ $\infty$ is not mathematically well-posed since such perfect equilibrium constraint is in general not achievable. Indeed, the demand at terminal date $T$ is random, typically modelled via a Gaussian noise, and the inventory $X$ which is of finite variation, may exceed or underperform with positive probability the demand $D_T$ at terminal date $T$. Hence, in the scenario where $X_T$ $>$ $D_T$, and since 
 by nature the production quantity $\xi$ is nonnegative, it is not possible to realize the equilibrium $X_T +\xi$ $=$ $D_T$, even if there is no delay.  In  the sequel, we fix 
 $\eta$ $>$ $0$ (which may be large, but finite), and study the stochastic control  problem \reff{defoptim}. 
 
\vspace{1mm} 

\noindent {\bf 4)} The optimization problem \reff{defoptim} shares somes similarities with the optimal execution problem in limit order book  studied in the seminal paper by Almgren and Chriss  \cite{almcri00}, and then extended by many authors in the literature, see e.g. the survey paper \cite{schsly11}. The main difference is that in the  execution problem of equities, the target is to buy or sell a  certain number of shares, i.e. lead $X_T$ to a fixed constant (meaning formally that $\eta$ goes to infinity)  while in our intraday electricity markets context, the target is to realize the equilibrium with the random demand $D_T$, eventually with the help of production leverage $\xi$.  However, in contrast with the case of constant target, it is not possible in presence of random target $D_T$ to achieve perfectly the equilibrium, which justifies the introduction of the penalty factor $\eta$ as pointed out above. 
\ep
}
\end{Remark}

 \vspace{1mm}

The main aim of this paper is to provide explicit  (or at least approximate explicit) solutions to the optimization problem \reff{defoptim}, which are easily  interpreted from an economic point of view, and also allow to measure the impact of the various parameters of the model. In order to achieve  this goal, we shall adapt our modeling as close as possible to the linear-quadratic framework of stochastic control, and make the following assumptions:  The energy  production  cost function is in the quadratic  form: 
\beqs
c(x) &=& \frac{\beta}{2} x^2, \;\;\;
\enqs
for some $\beta$ $>$ $0$.  Although simple, a quadratic cost function represents the increase of the marginal cost of production with the level of production.

\begin{Remark} \label{rembeta}  (\it{Pure retailer}) 
{\rm
In the limiting case  when  $\beta$ goes to infinity, meaning an infinite cost of production, this  
corresponds to the framework where the agent never uses the production leverage  and only trades in the intraday-market by solving the optimal execution problem:
\beq \label{optimnoprod}
 \mbox{ minimize over } q \in \Ac & & \E \Big[  \int_0^T q_t P_t(q) dt  + C(D_T-X_T,0) \Big].
\enq
\ep
}
\end{Remark}

As in Almgren and Chriss, we assume that the price impact (both permanent and temporary) is of linear form, i.e. 
\beqs
g(q) \; = \; \nu q, & & f(q) \; = \; \gamma q, 
\enqs
for some  constants $\nu$ $\geq$ $0$  and $\gamma$ $>$ $0$.  
The unaffected intraday electricity price is taken as a Bachelier model: 
\beq \label{Bachelier}
\hat P_t &=& \hat P_0 + \sigma_0 W_t, 
\enq
where $W$ is a standard Brownian motion, and $\sigma_0$ $>$ $0$ is a positive constant. Such assumption might seem a shortcoming at first sight since it allows for negative values of the unaffected price. However, in practice, for our intraday execution problem within few hours, negative prices occur only with negligible probability.  The martingale assumption is also standard in the market impact literature since drift effects can often be ignored due to short trading horizon.  
The quoted  price $Y$, impacted by the past trading rate $q$ $\in$ $\Ac$,  
is then governed by the dynamics: 
\beq \label{dynY}
dY_t &=&  \nu q_t dt + \sigma_0 dW_t. 
\enq 
The dynamics of the  residual demand forecast  is given by
\beq \label{dynD}
dD_t &=& \mu dt + \sigma_d dB_t, 
\enq
where $\mu$, $\sigma_d$ are constants, with $\sigma_d$ $>$ $0$,  and $B$ is a Brownian motion correlated with $W$: $d<W,B>_t$ $=$ $\rho dt$, $\rho$ $\in$ $[-1,1]$.

From  \reff{relPY}, one can then define the value function associated to the dynamic version of the optimal execution problem \reff{defoptim}  by:
\beq 
v(t,x,y,d) & :=&  \inf_{q \in \Ac_t,\xi \in L_+^0(\Fc_{T-h})} J(t,x,y,d;q,\xi) \label{defvpos}
\enq
with 
\beq \label{defJ}
J(t,x,y,d;q,\xi) &:=& \E \Big[ \int_t^T q_s (Y_s^{t,y} + \gamma q_s) ds + C(D_T^{t,d}- X_T^{t,x},\xi)^2   \Big],
\enq
for $(t,x,y,d)$ $\in$ $[0,T]\times\R\times\R\times\R$,  where  $\Ac_t$ denotes the set of real-valued processes 
$q$ $=$ $(q_s)_{t\leq s\leq T}$ s.t. $q_s$ is $\Fc_s$-adapted and $\E\big[\int_t^T q_s^2 ds]$ $<$ $\infty$, 
$D^{t,d}$ is the solution to \reff{dynD} starting from $d$ at  $t$, and  given a control $q$ $\in$ $\Ac_t$,  
$Y^{t,y}$ denotes the solution to \reff{dynY} starting from $y$ at  time $t$, and $X^{t,x}$ is the solution to \reff{dynX} starting from $x$ at $t$.

In a first step, we shall consider the case when there is no delay in the production, and then show in the last section of this paper how to reduce  the 
problem with delay to a no delay problem. We shall also study the case when  there are jumps in the residual demand forecast. 
 
\section{Optimal execution without delay in production}

\setcounter{equation}{0}
\setcounter{Assumption}{0} \setcounter{Theorem}{0}
\setcounter{Proposition}{0} \setcounter{Corollary}{0}
\setcounter{Lemma}{0} \setcounter{Definition}{0}
\setcounter{Remark}{0}

In this section, we consider the case when there is no delay in production, i.e. $h$ $=$ $0$.  
In this case, we notice that the optimization over $q$ and $\xi$ in \reff{defoptim} is  done separately. Indeed, the production quantity  $\xi$ 
$\in$ $L_+^0(\Fc_T)$ is chosen at the  final date $T$, after the decision over the trading rate process  $(q_t)_{t\in [0,T]}$ is achieved (leading to an inventory $X_T$).  It is  determined  optimally through the optimization a.s. at $T$ of the terminal cost $C(D_T-X_T,\xi)$, hence  in feedback form by 
$\xi_T^*$ $=$ $\hat\xi^+(D_T-X_T)$ where 
\beq
\hat\xi^+(d) & := & \argmin_{\xi \geq 0} C(d,\xi) 
\; = \; \argmin_{\xi \geq 0} \big[ \frac{\beta}{2} \xi^2  + \frac{\eta}{2} (d-  \xi)^2 \big] \nonumber \\
&=& \frac{\eta}{\eta+ \beta} d \ind_{d \geq 0}.  \label{xi+}
\enq
The  value function of problem \reff{defvpos} may then be rewritten as
\beq \label{defvpos2}
 v(t,x,y,d) & =& \inf_{q \in \Ac_t}  \E \Big[ \int_t^T q_s (Y_s^{t,y} + \gamma q_s) ds + C^+(D_T^{t,d}- X_T^{t,x})^2   \Big], 
\enq
where 
\beq  
C^+(d) &:=& C(d,\hat\xi^+(d))  \nonumber \\
& = &     \frac{1}{2} \frac{\eta \beta}{\eta + \beta} d^2 \ind_{d\geq 0} +  \frac{\eta}{2} d^2  \ind_{d< 0}.  \label{defC+}
\enq 
and the optimal trading rate $q^*$  is derived by solving \reff{defvpos2}.

 Due to the indicator function in $C^+$, caused by the non negativity constraint on the production quantity,  there is no hope to get explicit solutions for the problem \reff{defvpos2}, i.e. solve explicitly the associated dynamic programming Hamilton-Jacobi-Bellman (HJB)  equation. 
 We shall then consider an auxiliary  execution problem by relaxing  the sign constraint on the production quantity, for which we are able to provide explicit solution. Next, we shall see  how one can derive  an approximate solution to the original problem in terms of this auxiliary  explicit solution, and we evaluate the error and illustrate the quality of this approximation by numerical tests.

\subsection{Auxiliary optimal execution problem} \label{secaux}

We consider the optimal execution problem   with  relaxation on  the non negativity constraint  of the production leverage, and thus introduce the auxiliary value function 
\beq 
\tilde v(t,x,y,d) & :=&  \inf_{q \in \Ac,\xi \in L^0(\Fc_T)} J(t,x,y,d;q,\xi), \nonumber 
\enq
for $(t,x,y,d)$ $\in$ $[0,T]\times\R\times\R\times\R$. By same arguments as for the derivation of   \reff{defvpos2}, we have
\beq
\tilde v(t,x,y,d) &=&  \inf_{q \in \Ac} \E \Big[ \int_t^T q_s (Y_s^{t,y} + \gamma q_s) ds + \tilde C(D_T^{t,d}-X_T^{t,x}) \Big],  \label{defvaux}
\enq
where
\beq
\hat\xi(d)  & := & \argmin_{\xi \in \R} C(d,\xi)   \; = \;  \frac{\eta}{\eta + \beta} d, \nonumber \\
\tilde C(d) & :=& C(d,\hat\xi(d)) \; = \;  \frac{1}{2} \frac{\eta \beta}{\eta + \beta} d^2 \; = : \; \frac{1}{2} r(\eta,\beta) d^2. \label{deftildeC} 
\enq
The function in (\ref{deftildeC}) can be interpreted as a reduced cost function. Because the production cost function and the penalization are both quadratic, they can be reduced to a single production function where the imbalances are internalized by the producer.

By exploiting the linear-quadratic structure of the stochastic control problem \reff{defvaux}, we can obtain explicit solutions for this auxiliary problem.

\begin{Theorem} \label{thm1}
The value function to \reff{defvaux} is explicitly equal  to: 
\beqs
\tilde v(t,x,y,d) &=&  \frac{r(\eta,\beta)(\frac{\nu}{2}(T-t)+\gamma)}{ (r(\eta,\beta) +  \nu)(T-t)+2\gamma} \big( (d-x)^2+2\mu(T-t)(d-x)\big) \\
& & + \;  \frac{T-t}{ (r(\eta,\beta) +   \nu)(T-t)+2\gamma}  \big(-\frac{y^2}{2} + r(\eta,\beta)\mu(T-t)y\big) \\
& & + \;  \frac{r(\eta,\beta)(T-t)}{ (r(\eta,\beta) +   \nu)(T-t)+2\gamma} (d-x)y \\
& & + \; \gamma\frac{\sigma_0^2 + \sigma_d^2r^2(\eta,\beta) - 2\rho\sigma_0\sigma_dr(\eta,\beta)}{\big(r(\eta,\beta) +  \nu\big)^2} 
\ln\Big( 1 + \frac{(r(\eta,\beta) +  \nu)(T-t)}{2\gamma}\Big)  \\
& & + \; \frac{\sigma_d^2r(\eta,\beta)\nu+2\rho\sigma_0\sigma_dr(\eta,\beta) - \sigma_0^2}{2\big(r(\eta,\beta) + \nu\big)}(T-t) \\
& & + \;  \frac{r(\eta,\beta)\mu^2(T-t)^2(\frac{\nu}{2}(T-t)+\gamma)}{ (r(\eta,\beta) + \nu)(T-t)+2\gamma}, 
\enqs
for $(t,x,y,d)$ $\in$ $[0,T]\times\R\times\R\times\R$, with an optimal trading rate given in feedback form by:
\beq
\hat q_s &=&  \hat q\big(T-s,D_s^{t,d}-\hat X_s^{t,x,y,d},\hat Y_s^{t,x,y,d}\big), \;\;\; t\leq s\leq T  \nonumber \\
\hat q(t,d,y) &:=& \frac{ r(\eta,\beta) (\mu t + d) -   y}{(r(\eta,\beta) + \nu)t + 2 \gamma}. \label{optq} 
\enq
Here $(\hat X^{t,x,y,d},\hat Y^{t,x,y,d},D^{t,d})$ denotes the solution to \reff{dynX}-\reff{dynY}-\reff{dynD} when using the feedback control $\hat q$, and starting from $(x,y,d)$ at time $t$. Finally, the optimal  production leverage  is given by:
\beq \label{optxiaux}
\hat \xi_T &=& \hat\xi(D_T^{t,d}-\hat X_T^{t,x,y,d}) \; = \; \frac{\eta}{\eta + \beta} \big(D_T^{t,d} - \hat X_T^{t,x,y,d}\big). 
\enq
\end{Theorem}
{\bf Skech of proof.}  We look for a candidate solution to  \reff{defvaux}  in the quadratic form:
\beqs
\tilde w(t,x,y,d) &=& A(T-t) (d-x)^2 + B(T-t) y^2 + F(T-t) (d-x)y  \\
& & \;\;\; + \;  G(T-t) (d-x) + H(T-t) y + K(T-t),
\enqs 
for some deterministic functions $A$, $B$, $F$, $G$, $H$ and $K$. Plugging this ansatz into the Hamilton-Jacobi-Bellman (HJB) equation associated to the stochastic control problem \reff{defvaux}, we find that 
these deterministic functions should satisfy a system of Riccati equations, which can be explicitly solved. Then, by  a classical verification argument, we check that this ansatz $\tilde w$ is indeed equal to $\tilde v$, with an optimal feedback control derived from the argmax in the HJB equation. The details of the proof are reported in Appendix. 
\ep

\vspace{5mm}

\begin{Remark} \label{rembeta2}
{\rm  ({\em Pure trader}) By sending $\beta$ to infinity in the expression of the value function $\tilde v$ and of the optimal feedback control $\hat q$, and observing that $r(\eta,\beta)$ goes to $\eta$, we obtain the solution to the optimal execution problem \reff{optimnoprod} without leverage production: 
\beq
v_{_{NP}}(t,x,y,d) & :=&   \inf_{q \in \Ac} \E \Big[ \int_t^T q_s (Y_s^{t,y} + \gamma q_s) ds + C(D_T^{t,d}-X_T^{t,x},0) \Big] \label{defw0} \\
&=&  \frac{\eta(\frac{\nu}{2}(T-t)+\gamma)}{ (\eta  +  \nu)(T-t)+2\gamma} \big( (d-x)^2+2\mu(T-t)(d-x)\big) \nonumber \\
& & + \;  \frac{T-t}{ (\eta  +  \nu)(T-t)+2\gamma}  \big(-\frac{y^2}{2} + \eta\mu(T-t)y\big) \nonumber \\
& & + \;  \frac{\eta(T-t)}{ (\eta  +  \nu)(T-t)+2\gamma} (d-x)y \nonumber \\
& & + \; \gamma\frac{\sigma_0^2 + \sigma_d^2\eta^2 - 2\rho\sigma_0\sigma_d\eta}{\big(\eta +\nu\big)^2} 
\ln\Big( 1 + \frac{(\eta +  \nu)(T-t)}{2\gamma}\Big)  \nonumber \\
& & + \; \frac{\sigma_d^2\eta\nu+2\rho\sigma_0\sigma_d\eta - \sigma_0^2}{2\big(\eta + \nu\big)}(T-t) \nonumber \\
& & + \;  \frac{\eta\mu^2(T-t)^2(\frac{\nu}{2}(T-t)+\gamma)}{ (\eta +  \nu)(T-t)+2\gamma}, \nonumber
\enq
for $(t,x,y,d)$ $\in$ $[0,T]\times\R\times\R\times\R$, with an optimal trading rate given in feedback form by:
\beqs
\label{eq:qnojump}
\hat q_s^{NP} &=&  \hat q^{NP}\big(T-s,D_s^{t,d}-\hat X_s^{t,x,y,d},\hat Y_s^{t,x,y,d}\big), \;\;\; t\leq s\leq T  \\
\hat q^{NP}(t,d,y) &:=& \frac{ \eta (\mu t + d) -   y}{(\eta + \nu)t + 2 \gamma}. \nonumber
\enqs
\ep
}
\end{Remark}

\vspace{5mm}

\noindent {\bf Interpretation:} 
\begin{itemize}
\item[1.] The optimal trading rate $\hat q_s$ at time $s$ $\in$ $[t,T]$, given in feedback form by \reff{optq}, is decomposed in two terms:  the first one
\beqs
 \frac{ r(\eta,\beta)} {(r(\eta,\beta) +  \nu)(T-t) + 2 \gamma} \big( \mu(T-s) + D_s^{t,d} -\hat X_s^{t,x,y,d}\big)
\enqs
is related to the trading rate  in order to follow the trend of the demand, and to the incentive to invest when the forecast of the residual demand is larger than the current inventory. The second term 
\beqs
-  \frac{1} {(r(\eta,\beta) + \nu)(T-t) + 2 \gamma}  \hat Y_s^{t,x,y,d}
\enqs
represents the negative impact of the quoted  price on the investment strategy: the higher the price is, the more the agent decreases her trading rate until she reaches negative value meaning 
a resale of electricity shares.   These effects are weighted by  the constant denominator term  depending on the penalty factor $\eta$, 
the marginal  cost production factor $\beta$, the temporary and permanent price impact  parameters $\gamma$, $\nu$, and the time to maturity $T-t$.

\item[2.]  By introducing the marginal cost function: $c'(x)$ $=$ $\beta x$, and the process 
\beqs
\hat\xi_s & := & \frac{\eta}{\eta + \beta} \big( D_s^{t,d} + \mu(T-s) - \hat X_s^{t,x,y,d} - \hat q_s  (T-s) \big), \;\; t \leq s \leq T,
\enqs
which is interpreted as the forecast production for the final time $T$ (recall expression \reff{optxiaux} of the final production), we notice from the expression of the 
optimal trading rate that the following relation holds:
\beq
\label{eq:qnojump2}
\hat Y_s^{t,x,y,d}  + \nu \hat q_s  (T-s)  + 2 \gamma \hat q_s &=&   c'(\hat\xi_s),  \;\;\; t \leq s \leq T.
\enq
This relation means that at each time, the optimal trading rate is to make the forecast intraday price plus marginal temporary impact (left hand side), which can be seen as the marginal cost of electricity on the intraday market at time $T$, equal to the forecast marginal cost of production.  
Here, the instantaneous impact $\gamma$ appears  as a marginal cost of buying or selling, and the forecast at time $s$  
supposes that  the optimal trading rate $\hat q_s$ is held constant between $s$ and $T$.
\ep
\end{itemize}

\vspace{3mm}

We complete the description of the optimal trading rate  by  pointing out a remarkable martingale property.

\begin{Proposition} \label{propmartinq}
The optimal trading rate process $(\hat q_s)_{t\leq s \leq T}$ in \reff{optq} is a martingale. 
\end{Proposition}
{\bf Proof.} By applying It\^o's formula to $\hat q_s$ $=$ $\hat q(T-s,D_s^{t,d}-\hat X_s^{t,x,y,d},\hat Y_s^{t,x,y,d})$, $t\leq s\leq T$, and since $\hat q$ is linear in $d$ and $y$, we have:
\beqs
d\hat q_s &=& \big[ - \Dt{\hat q} + (\mu - \hat q)\Dd{\hat q} + \nu \hat q \Dy{\hat q} \big]
(T-s,D_s^{t,d}-\hat X_s^{t,x,y,d},\hat Y_s^{t,x,y,d}) ds \\
& & \; + \; \Dd{\hat q}(T-s,D_s^{t,d}-\hat X_s^{t,x,y,d},\hat Y_s^{t,x,y,d})  \sigma_d dB_s \\
& & \; + \;   \Dy{\hat q}(T-s,D_s^{t,d}-\hat X_s^{t,x,y,d},\hat Y_s^{t,x,y,d}) \sigma_0 dW_s, 
\enqs
from the dynamics \reff{dynX}, \reff{dynD}, and \reff{dynY} of  $\hat X^{t,x,y,d}$, $D^{t,d}$ and $\hat Y^{t,x,y,d}$. Now, from the explicit expression of the  function $\hat q(t,y,d)$, we see that
\beqs
- \Dt{\hat q} + (\mu - \hat q) \Dd{\hat q} + \nu \hat q \Dy{\hat q} &=& 0,
\enqs
and so: 
\beq \label{hatqmar}
d\hat q_s &=& 
\frac{r(\eta,\beta)\sigma_d }{(r(\eta,\beta) + \nu)(T-s) + 2 \gamma} dB_s -   \frac{\sigma_0 }{(r(\eta,\beta) + \nu)(T-s) + 2 \gamma} dW_s,
\enq
which shows the required martingale property. 
\ep

\begin{Remark} \label{remmartinq}
{\rm Recall that in the classical optimal execution problem as studied in \cite{almcri00}, the optimal trading rate is constant. We retrieve this result in their framework which corres\-ponds to the case where $\sigma_d$ $=$ $0$ (constant demand target),  $\beta$ $=$ $\infty$ (there is no production), and $\eta$ $=$ $\infty$ (constraint to lead $X_T$ to the fixed target), see Remark \ref{rem1} {\bf 4)}. Indeed, in these limiting regimes, we see from \reff{hatqmar} that $d\hat q_s$ $=$ $0$, meaning that $\{\hat q_s,t\leq s\leq T\}$ is constant.  In our framework, this is generalized to the martingale property of the optimal trading rate process, which implies that the optimal inventory $\{\hat X_s^{t,x,y,d}$, $t\leq s\leq T\}$ has a constant growth rate in mean, i.e.  $\frac{d \E[\hat X_s^{t,x,y,d}]}{ds}$  is constant equal to the initial trading rate at time $t$ given by $\hat q(T-t,d-x,y)$.  

As a  consequence of this martingale property,  if the producer already satisfies the relation \reff{eq:qnojump2}  in the day-ahead market, and if the initial intraday price is the day-ahead price, her initial trading rate on the intraday market will be zero. And thus, on average, her trading rate will be zero. 

The martingale property of the trading rate process is actually closely related to the martingale dynamics of the unaffected price $\hat P$ in \reff{Bachelier}.  As we shall see in Section~\ref{secjump} where we consider jumps on price, making $\hat P$ a sub or super martingale, the optimal trading rate will inherit the converse  sub or super martingale property. 
\ep
}
\end{Remark}

\subsection{Approximate solution} \label{secapprox}

 We go back to the original execution problem with the  non negativity constraint on the production quantity. As pointed out above, there is no explicit solution in this case, due to the form of the terminal cost function $C^+$. The strategy  is then to use the explicit  control consisting in  the trading rate $\hat q$ derived in \reff{optq}, and of the truncated nonnegative production quantity: 
 \beq \label{tildexi} 
 \tilde\xi_T^* & := & \hat \xi_T \ind_{_{\hat\xi_T \geq 0}} \; = \;  \hat\xi^+(D_T^{t,d} - \hat X_T^{t,x,y,d}),
 \enq
 with $\hat\xi_T$ defined in \reff{optxiaux} from the auxiliary problem. In other words, we follow the  trading rate strategy  $\hat q$ determined from the problem without constraint on the final production quantity, and at the terminal date use the production leverage if  the final inventory 
 $\hat X_T^{t,x,y,d}$  is below the terminal demand $D_T^{t,d}$, by choosing  a quantity proportional to this spread $D_T^{t,d}-\hat X_T^{t,x,y,d}$.   
 The aim of this section is to measure the relevance of this approximate strategy 
 $(\hat q,\tilde\xi_T^*)$ $\in$ $\Ac\times L_+^0(\Fc_T)$ with respect to the optimal execution problem \reff{defvpos} by estimating the  induced error:
 \beqs
\Ec_1(t,x,y,d) & := &  J(t,x,y,d;\hat q,\tilde\xi_T^*) - v(t,x,y,d),
 \enqs
for $(t,x,y,d)$ $\in$ $[0,T]\times\R\times\R\times\R$.  We also measure the approximation error on the value functions:
\beqs
\Ec_2(t,x,y,d) & := & v(t,x,y,d) - \tilde v(t,x,y,d).
\enqs
Notice that if $\hat\xi_T$ $\geq$ $0$ a.s., i.e. $D_T^{t,d}$ $\geq$ $\hat X_T^{t,x,y,d}$ a.s. (which is not true), and so $\tilde\xi_T^*$ $=$ $\hat\xi_T$, then clearly $(\hat q,\hat\xi_T)$  would be the solution to \reff{defvpos}, and so $\Ec_1(t,x,y,d)$ $=$ $\Ec_2(t,x,y,d)$ $=$ $0$. Actually, these errors depend on the probability of the event: $\{\hat X_T^{t,x,y,d}$ $>$ $D_T^{t,d}\}$, and we have the following estimate: 
 
\begin{Proposition} \label{erreurnopanne}
For all $(t,x,y,d)$ $\in$ $[0,T]\times\R\times\R\times\R$, we have
\beq \label{estimerror}
0 \; \leq \; \Ec_i(t,x,y,d)  & \leq & \frac{\eta r(\eta,\beta)}{2\beta} V(T-t) \psi \Big( \frac{m(T-t,d-x,y)}{\sqrt{V(T-t)}}\Big), \;\;\;  i=1,2, 
\enq
where 
\beqs
\psi(z) &:=& (z^2 + 1) \Phi(-z) - z\phi(z), \;\;\; z \in \R,
\enqs
with $\phi$ $=$ $\Phi'$ the density of the standard normal distribution,  and
\beq
m(t,d,y) &:=& \frac{(\nu t+2\gamma)(\mu t+d) +   y t}
{(r(\eta,\beta) +  \nu)t + 2 \gamma}, \label{defm} \\
V(t) &:=& \int_0^t \frac{\sigma_0^2 s^2 + \sigma_d^2(\nu s + 2 \gamma)^2 + 2 \rho\sigma_0\sigma_ds (\nu s + 2 \gamma)} 
{\big[(r(\eta,\beta) + \nu)s + 2 \gamma\big]^2} ds \; \geq \; 0. \label{defVar}
\enq
\end{Proposition} 
{\bf Proof.} By definition of the value functions $v$ and $\tilde v$, recalling that $(\hat q,\hat\xi_T)$ is an optimal control for $\tilde v$,  
and since $(\hat q,\tilde\xi_T^*)$ $\in$ $\Ac\times L_+^0(\Fc_T)$, we have: 
\beqs
J(t,x,y,d;\hat q,\hat\xi_T) \; = \; \tilde v(t,x,y,d) \; \leq \;  v(t,x,y,d) & \leq &  J(t,x,y,d;\hat q,\tilde\xi_T^*),
\enqs
for all $(t,x,y,d)$ $\in$ $[0,T]\times\R\times\R\times\R$. This clearly implies that both errors $\Ec_1$ and $\Ec_2$ are nonnegative, and 
\beqs
\max(\Ec_1(t,x,y,d),\Ec_2(t,x,y,d) ) & \leq & \Ec(t,x,y,d) \; := \; J(t,x,y,d;\hat q,\tilde\xi_T^*) - J(t,x,y,d;\hat q,\hat\xi_T). 
\enqs
We now focus on the upper bound for  $\Ec$. By definition of $J$ in \reff{defJ},  $\hat\xi_T$ and $\tilde\xi_T^*$ in \reff{optxiaux} and \reff{tildexi}, we have
\beq
\Ec(t,x,y,d) & =& \E \Big[ C(D_T^{t,d} - \hat X_T^{t,x,y,d},\tilde\xi_T^*) - C(D_T^{t,d} - \hat X_T^{t,x,y,d},\hat\xi_T) \Big] \nonumber \\
&=&   \E \Big[ C(D_T^{t,d} - \hat X_T^{t,x,y,d},  \hat\xi^+(D_T^{t,d} - \hat X_T^{t,x,y,d}) )  \nonumber \\
& & \hspace{2cm} - \;  C(D_T^{t,d} - \hat X_T^{t,x,y,d},\hat\xi(D_T^{t,d} - \hat X_T^{t,x,y,d})) \Big]  \nonumber \\
&=&   \E \Big[ C^+(D_T^{t,d} - \hat X_T^{t,x,y,d} ) -  \tilde C(D_T^{t,d} - \hat X_T^{t,x,y,d} ) \Big] \nonumber \\
&=& \frac{\eta r(\eta,\beta)}{2\beta} \E \Big[  \big(D_T^{t,d} - \hat X_T^{t,x,y,d} \big)^2 \ind_{D_T^{t,d} - \hat X_T^{t,x,y,d} < 0} \Big],  \label{expressinterEc}
\enq
from the definitions and expressions of $C^+$ and $\tilde C$ in \reff{defC}, \reff{defC+} and \reff{deftildeC}. Now, from \reff{hatqmar} and  
by integration,  we obtain the explicit (path-dependent) form of the optimal trading rate control:
\beqs
\hat q_s &=&  \hat q_t 
+ \int_t^s \frac{r(\eta,\beta) \sigma_d} {(r(\eta,\beta) + \nu)(T-u) + 2 \gamma}  dB_u  \\
& & \;\;\; - \;   \int_t^s \frac{ \sigma_0} {(r(\eta,\beta) + \nu)(T-u) + 2 \gamma}  dW_u, \;\;\; t \leq s \leq T,
\enqs
with $\hat q_t$ $=$ $\hat q(T-t,d-x,y)$. We then obtain the expression of the final spread between demand and inventory: 
\beqs
D_T^{t,d} - \hat X_T^{t,x,y,d} &=& d- x +  \mu(T-t) +  \int_t^T  \sigma_d dB_s -   \int_t^T \hat q_s ds  \\
&= & m(T-t,d-x,y)  \; + \int_t^T \frac{\sigma_d(\nu(T-s)+2\gamma)}{(r(\eta,\beta) + \nu)(T-s) + 2 \gamma} dB_s \\
& & \; + \; \int_t^T \frac{\sigma_0 (T-s)}{(r(\eta,\beta) + \nu)(T-s) + 2 \gamma} dW_s,  
\enqs 
by Fubini's theorem, and with
\beqs
m(t,d,y) & := &  d + \mu t   - t \hat q(t,d,y), 
\enqs
which is explicitly written as in \reff{defm} from the expression \reff{optq} of $\hat q$.  Thus, $D_T^{t,d} - \hat X_T^{t,x,y,d}$ follows a normal distribution law  with mean $m(T-t,d-x,y)$ and variance $V(T-t)$ given by \reff{defVar}, and from \reff{expressinterEc}, we deduce that
\beqs
\Ec(t,x,y,d) & = & \frac{\eta r(\eta,\beta)}{2\beta}  V(T-t) \psi \Big( \frac{m(T-t,d-x,y)}{\sqrt{V(T-t)}}\Big), 
\enqs 
while the probability that the final inventory is larger than the terminal demand is:
\beq \label{X<D}
\P\big[ D_T^{t,d} - \hat X_T^{t,x,y,d} < 0 \big] &=&  \Phi \Big( - \frac{m(T-t,d-x,y)}{\sqrt{V(T-t)}}\Big). 
\enq
\ep

\vspace{5mm}

\noindent {\bf Error asymptotics.}  We now investigate the accuracy of the upper bound in \reff{estimerror}
\beqs
\bar\Ec(T-t,d-x,y) &:=& \frac{\eta r(\eta,\beta)}{2\beta} V(T-t) \psi \Big( \frac{m(T-t,d-x,y)}{\sqrt{V(T-t)}}\Big).
\enqs 
It is well-known (see e.g. Section~14.8 in \cite{wil91}) that 
\beq \label{relphi}
 z \Phi(-z) & \leq & \phi(z), \;\;\; \forall z \in \R, 
\enq
from which we easily see that $\psi$ is non increasing, convex, and $\psi(\infty)$ $=$ $0$. Thus, $\bar\Ec(T-t,d-x,y)$ decreases to zero for large $m(T-t,d-x;y)$ or small $V(T-t)$.  
We shall study its asymptotics in three limiting cases (i) the time to maturity $T-t$ is small,  (ii)  the initial demand spread  $d-x$ is large, (iii) the initial quoted price $y$ is large.  We prove that 
the error bound $\bar\Ec(T-t,d-x,y)$, and thus $\Ec_1(t,x,y,d)$, $\Ec_2(t,x,y,d)$, converge  to zero at least with an exponential rate of convergence in these limiting regimes:

\begin{Proposition} \label{properrorlimite}
\begin{itemize}
\item[(i)] For all  $(x,y,d)$ $\in$ $\R\times\R\times\R$ with $d$ $>$ $x$, we have
\beq \label{estimT-t}
\limsup_{T-t\downarrow 0} (T-t) \ln  \bar\Ec(T-t,d-x,y) & \leq & - \frac{1}{2} \Big(\frac{d-x}{\sigma_d} \Big)^2.
\enq
\item[(ii)] For all $(t,y)$ $\in$ $[0,T)\times\R$, we have
\beq \label{estimd-x}
\limsup_{d-x\rightarrow \infty} \frac{1}{(d-x)^2}  \ln  \bar\Ec(T-t,d-x,y) & \leq & - \frac{1}{2} \frac{m_\infty^2(T-t)}{V(T-t)},
\enq
where 
\beqs
m_\infty(t) &=&  \frac{\nu t+2\gamma}{\big( r(\eta,\beta) + \nu \big)t + 2 \gamma}
\enqs
\item[(iii)] For all $(t,x,d)$ $\in$ $[0,T)\times\R\times\R$, we have
\beq \label{estimy}
\limsup_{ y \rightarrow \infty} \frac{1}{y^2}  \ln  \bar\Ec(T-t,d-x,y) & \leq & - \frac{1}{2} \frac{n_\infty^2 (T-t)}{V(T-t)},
\enq
where 
\beqs
n_\infty(t) &=&  \frac{t}{(r(\eta,\beta) +  \nu)t + 2 \gamma}.
\enqs
\end{itemize}
\end{Proposition}
{\bf Proof.}
From \reff{relphi}, we have: 
\beqs
0 \;  \leq \; & \psi(z) & \leq \; z^{-1} \phi(z), \;\;\; \forall z >0. 
\enqs
Notice that in the three  asymptotic regimes (i) (with $d-x$ $>$ $0$), (ii), and (iii), the quantity $m(T-t,d-x,y)$ is positive, and we thus have:
\beq \label{estimbarEc}
\bar\Ec(T-t,d-x,y) & \leq & \frac{\eta r(\eta,\beta)}{2\beta}  \frac{V(T-t)^{\frac{3}{2}}}{m(T-t,d-x,y)} \phi \Big( \frac{m(T-t,d-x,y)}{\sqrt{V(T-t)}}\Big). 
\enq
(i) For small time to maturity $T-t$, we see that $m(T-t,d-x,y)$ converges to $d-x$ $>$ $0$, while  $V(T-t)$ $\sim$ $\sigma_d^2(T-t)$, i.e. 
$V(T-t)/\sigma^2_d(T-t)$ converges to $1$.  This shows from \reff{estimbarEc} that, when $T-t$ goes to zero, the error bound 
$\bar\Ec(T-t,d-x,y)$,  converges to zero at least with an exponential rate of convergence, namely the one given by \reff{estimT-t}.

\vspace{1mm}

\noindent (ii) For large demand spread $d-x$, we see that $m(T-t,d-x,y)$ $\sim$  $m_\infty(T-t) (d-x)$, i.e. the ratio $m(T-t,d-x,y)/m_\infty(T-t)(d-x)$ converges to 
$1$ when $d-x$ goes to infinity.  This shows from \reff{estimbarEc} that, when $d-x$ goes to infinity,   the error bound 
$\bar\Ec(T-t,d-x,y)$, converges to zero at least with an exponential rate of convergence, namely the one given by \reff{estimd-x}.
\vspace{1mm}

\noindent (iii) For large $y$, we see that $m(T-t,d-x,y)$ $\sim$  $n_\infty(T-t) y$, i.e. the ratio $m(T-t,d-x,y)/n_\infty(T-t)y$ converges to 
$1$ when $y$ goes to infinity. This shows from \reff{estimbarEc} that, when $d-x$ goes to infinity,   the error bound 
$\bar\Ec(T-t,d-x,y)$ converges to zero at least with an exponential rate of convergence, namely the one given by \reff{estimy}.
\ep

\vspace{2mm}

\noindent {\bf Interpretation}.  Recall from \reff{X<D} that 
\beqs 
\P\big[ D_T^{t,d}  <  \hat X_T^{t,x,y,d} \big] &=&  \Phi \Big( -  \frac{m(T-t,d-x,y)}{\sqrt{V(T-t)}}\Big), 
\enqs
and thus  following the same arguments as  in the above proof, we have:  
\begin{itemize}
\item[(i)]  
\beq \label{estim2T-t}
\limsup_{T-t\downarrow 0} (T-t) \ln  \P\big[ D_T^{t,d}  <  \hat X_T^{t,x,y,d} \big] & = & - \frac{1}{2} \Big(\frac{d-x}{\sigma_d} \Big)^2, 
\enq
for all  $(x,y,d)$ $\in$ $\R\times\R\times\R$ with $d$ $>$ $x$. 
We observe that the rate in the rhs of \reff{estimT-t} (or \reff{estim2T-t}) depends only on the demand volatility $\sigma_d$ and the initial demand spread $d-x$. 
Moreover, it is  all the larger,  the smaller  $\sigma_d$ is, and the larger $d-x$ is.  This means  that the terminal demand will stay with very high probability above the final inventory once we are near from the maturity with a low volatile demand, initially larger than the inventory, in which case, the  explicit strategy $(\hat q,\tilde\xi_T^*)$ approximates very accurately the optimal strategy  $(q^*,\xi_T^*)$. 
 \item[(ii)] 
\beq \label{estim2d-x}
\limsup_{d-x\rightarrow \infty} \frac{1}{(d-x)^2}  \ln   \P\big[ D_T^{t,d}  <  \hat X_T^{t,x,y,d} \big]  & = & - \frac{1}{2} \frac{m_\infty^2(T-t)}{V(T-t)},
\enq
for all $(t,y)$ $\in$ $[0,T)\times\R$,  The rate in the rhs of \reff{estimd-x} (or \reff{estim2d-x}) is all the larger, the smaller  the volatilities $\sigma_0$ and $\sigma_d$ of the electricity price and demand are.  Again, we have the same interpretation than in the asymptotic regime (i), and this means that the 
explicit strategy  $(\hat q,\tilde\xi_T^*)$ approximates very accurately the optimal strategy $(q^*,\xi_T^*)$ in the limiting regime when the initial demand spread is large, and the volatilities are small. 
\item[(iii)] 
\beq \label{estim2y}
\limsup_{ y \rightarrow \infty} \frac{1}{y^2}  \ln   \P\big[ D_T^{t,d}  <  \hat X_T^{t,x,y,d} \big]   & = & - \frac{1}{2} \frac{n_\infty^2 (T-t)}{V(T-t)},
\enq
for all $(t,x,d)$ $\in$ $[0,T)\times\R\times\R$. In the limiting regime where the initial quoted price $y$ is large,  the agent has a strong incentive to sell energy on the intraday market, which leads to a final inventory staying under the final demand with 
high probability, and thus to a very accurate approximate strategy  $(\hat q,\tilde\xi_T^*)$.  As in case (ii), this accuracy is strengthened for small volatilities $\sigma_0$ and $\sigma_d$ of the electricity price and demand. 
\ep
\end{itemize}

\subsection{Numerical results}

 \subsubsection{Numerical tests}

We measure quantitatively the accuracy of the error bound derived in the previous paragraph with some numerical tests. Let us fix the following parameter values: 
$\sigma_0=1/60$ \euro{}$\cdot$(MW)$^{-1}\cdot s^{-1/2}$, $\sigma_d=1000/60$ MW$\cdot s^{-1/2}$, $\beta=0.002$ \euro{}$\cdot$(MW)$^{-2}$, $
\eta=200$ \euro{}$\cdot$(MW)$^{-2}$,  
$\mu=0$ MW$\cdot s^{-1}$, 
$\nu=10^{-10}$\euro{}$\cdot$(MW)$^{-2}$, $\gamma=10^{-10}$\euro{}$\cdot s\cdot$(MW)$^{-2}$ and  $\rho=0.8$.
 
We start from the initial time $t$ $=$ $0$, with a zero inventory $X_0$ $=$ $0$, and vary respectively the maturity $T$, the initial demand $D_0$ and the initial price $Y_0$. We  compute the 
probability for the final inventory to exceed the final demand $\P[\hat X_T > D_T]$, the approximate value function $\tilde v(0,X_0,Y_0,D_0)$, and the error bound $\bar\Ec(T,D_0-X_0,Y_0)$. 
The results are reported in Table \ref{tab:testsHorizonTemporelVariant} when varying $T$, in Table \ref{tab:testsDemandeInitialeVariante}  when varying $D_0$ and in Table \ref{tab:testsPrixInitialVariant} when varying $Y_0$.

\begin{table}[h]
        \centering
        \begin{tabular}{|c|c|c|c|} \hline
                 $T$ (h) & $\P[\hat X_T> D_T]$  &  $\tilde v(0,X_0,Y_0,D_0)$ (\euro{}) & $\bar\Ec(T,D_0-X_0,Y_0)$ (\euro{}) \\
                \hline
                $1$ & $<10^{-16}$ & $1.88\times 10^6$ & $<10^{-16}$ \\
                $8$ & $<10^{-16}$ & $1.88\times 10^6$ & $<10^{-16}$ \\
                $24$ & $<10^{-16}$ & $1.89\times 10^6$ & $4.16\times 10^{-12}$ \\
                $50$ & $7.72\times 10^{-13}$ & $1.90\times 10^6$ & $2.48\times 10^{-4}$ \\
                \hline
        \end{tabular}
        \caption{$Y_0=50$ \euro{}$\cdot$(MW)$^{-1}$ and  $D_0=50000$ MW}
        \label{tab:testsHorizonTemporelVariant}
\end{table}

Table 1 shows that for time to maturity less than $T$ $=$ $24$h,  the probability for the final inventory to exceed the final demand is very small, and consequently the error bound is rather negligible. 
When the time horizon increases, the agent has the possibility to spread over time her trading strategies for reducing the price impact, and purchase more energy, in which case the probability for the final inventory to exceed the demand increases.

\vspace{2mm}

\begin{table}[h]
        \centering
        \begin{tabular}{|c|c|c|c|} \hline
                  $D_0$ (MW) & $\P[\hat X_T> D_T]$  & $\tilde v(0,X_0,Y_0,D_0)$ (\euro{}) & $\bar\Ec(T,D_0-X_0,Y_0)$ (\euro{}) \\
                \hline
                $500$ & $<10^{-16}$ & $-5.86\times 10^5$ & $4.16\times 10^{-12}$ \\
                $5000$ & $<10^{-16}$ & $-3.62\times 10^5$ & $4.16\times 10^{-12}$ \\
                $50000$ & $<10^{-16}$ & $1.89\times 10^6$ & $4.16\times 10^{-12}$ \\
                $500000$ & $<10^{-16}$ & $2.44\times 10^7$ & $4.16\times 10^{-12}$ \\
                \hline
        \end{tabular}
        \caption{$T=24$ h and  $Y_0=50$ \euro{}$\cdot$(MW)$^{-1}$}
        \label{tab:testsDemandeInitialeVariante}
\end{table}
 
Table 2 shows that the probability for the final inventory to exceed the final demand, and the error bound are  not much sensitive to the variations of the initial positive demand $D_0$. Actually, the main impact is caused by the initial stock price, as observed in Table 3.

 \vspace{2mm}

\begin{table}[h]
        \centering
        \begin{tabular}{|c|c|c|c|} \hline
                 $Y_0$ (\euro{}$\cdot$(MW)$^{-1}$) & $\P[\hat X_T> D_T]$ & $\tilde v(0,X_0,Y_0,D_0)$(\euro{}) & $\bar\Ec(T,D_0-X_0,Y_0)$ (\euro{}) \\
                \hline
                $500$ & $<10^{-16}$ & $2.51\times 10^6$ & $<10^{-16}$ \\
                $50$ & $<10^{-16}$ & $1.89\times 10^6$ & $4.16\times 10^{-12}$ \\
                $40$ & $9.51\times 10^{-15}$ & $1.61\times 10^6$ & $3.80\times 10^{-4}$ \\
                $30$ & $4.57\times 10^{-10}$ & $1.29\times 10^6$ & $1.30\times 10^{-2}$ \\
                $20$ & $2.23\times 10^{-5}$ & $9.13\times 10^5$ & $1.26\times 10^{3}$ \\
                \hline
        \end{tabular}
        \caption{$T=24$ h and  $D_0=50000$ MW}
        \label{tab:testsPrixInitialVariant}
\end{table}

For small initial electricity price $Y_0$, the agent will buy more energy in the intraday market and produce less. Therefore, the  inventory will overtake with higher probability the demand,  in which case the approximate value function can be significantly different from the original one, as  observed from the error bound in Table 3 for $Y_0$ $=$  $20$.

\subsubsection{Simulations} \label{simul}

We plot trajectories of some relevant quantities that we simulate with the following set of parameters:  $\sigma_0=1/60$ \euro{}$\cdot$(MW)$^{-1}\cdot s^{-1/2}$, $\sigma_d=1000/60$ MW$\cdot s^{-1/2}$, 
$\beta=0.002$ \euro{}$\cdot$(MW)$^{-2}$, $\eta=100$ \euro{}$\cdot$(MW)$^{-2}$,  $\mu=0$ MW$\cdot s^{-1}$,  $\rho=0.8$, 
$\nu=4.00\times 10^{-5}$\euro{}$\cdot$(MW)$^{-2}$, $\gamma=2.22$\euro{}$\cdot s\cdot$(MW)$^{-2}$,  $T$ $=$ $24$h, $X_0$ $=$ $0$, $D_0=50000$ MW and  $Y_0=50$ \euro{}$\cdot$(MW)$^{-1}$. 

For such parameter values, the probability $\P[\hat X_T > D_T]$ is bounded above by $10^{-16}$,  the error $\bar\Ec(0,D_0 -X_0,Y_0)$ is bounded by $2.82\times 10^{-10}$\euro{}, and 
\beqs
\tilde v(0,X_0,Y_0,D_0) &=& 1916700 \text{\euro{}.}
\enqs
The executed strategy $(\hat q,\hat\xi_T^*)$ can then be considered as very close to the optimal strategy. 

\vspace{2mm}

\begin{figure}[h]
        \centering
        \includegraphics[height=5cm,width=\textwidth]{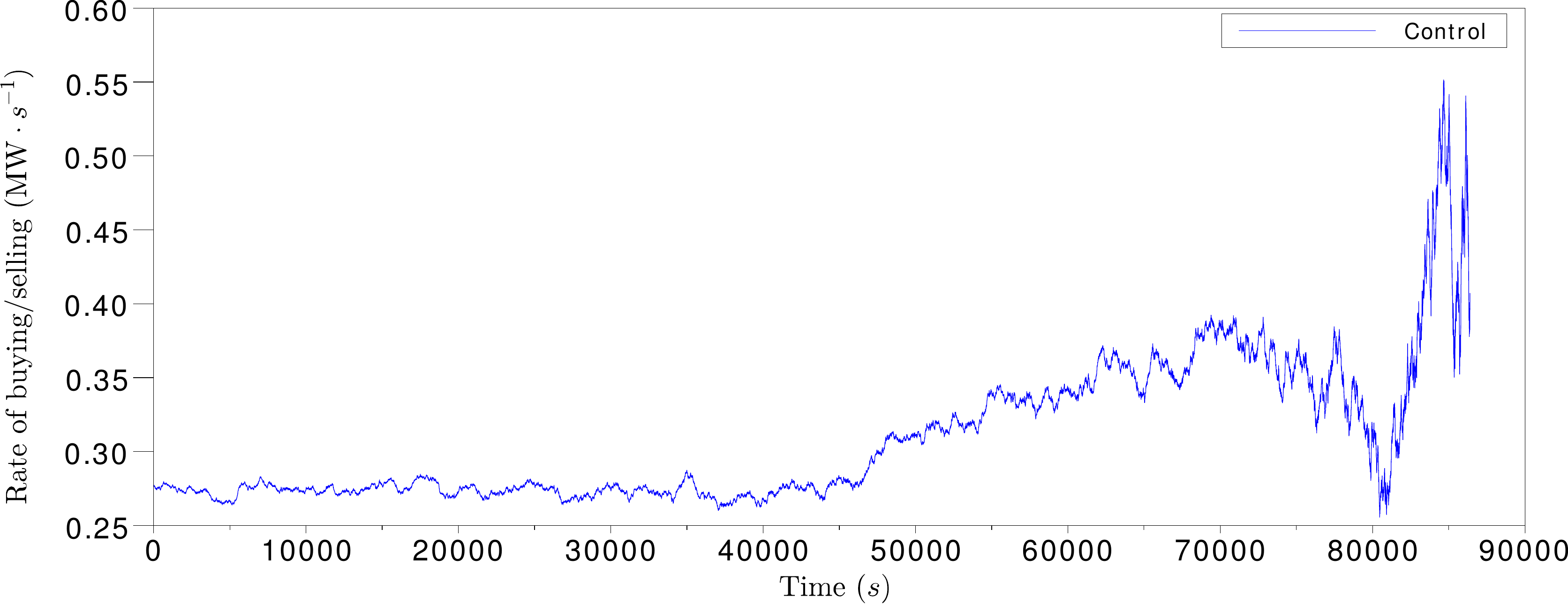}
        \caption{\small{Evolution of the  trading rate control $\hat q$}}
        \label{fig:evolutionControlePb}
\end{figure}

\begin{figure}[h]
        \centering
        \includegraphics[height=5cm,width=\textwidth]{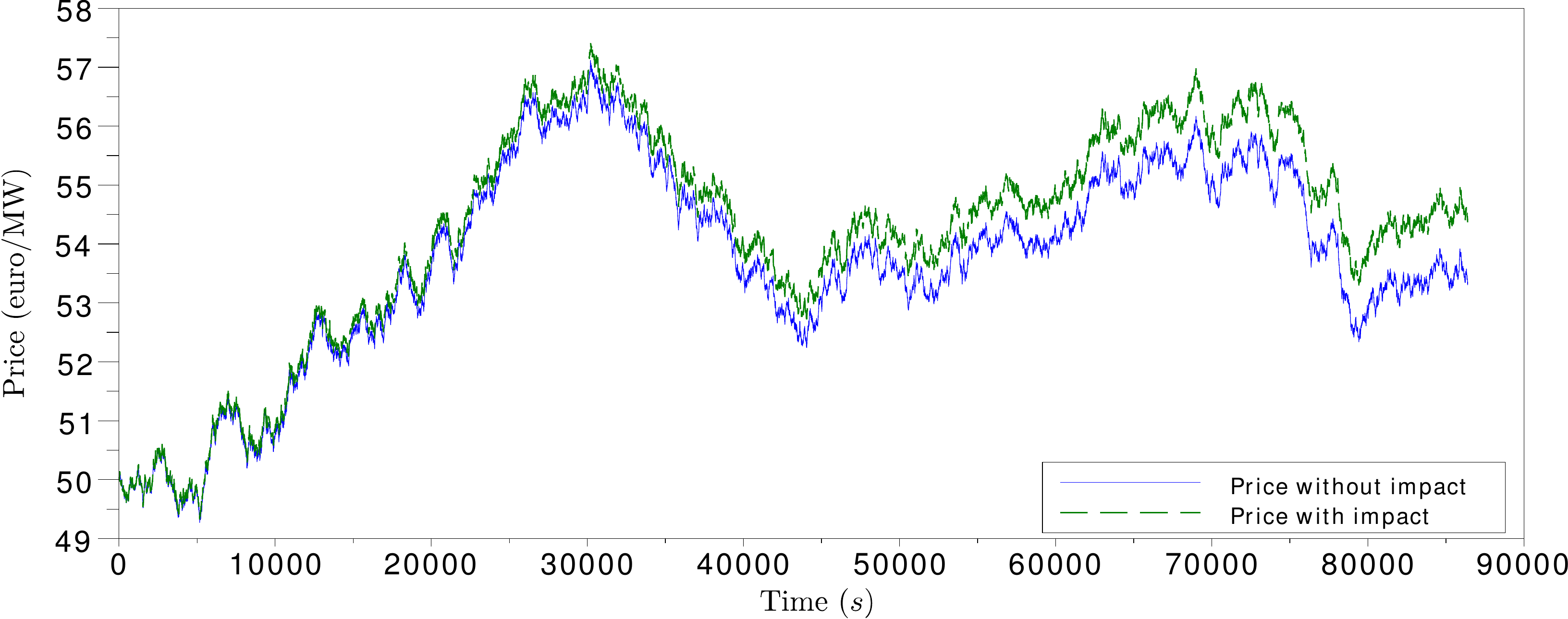}
        \caption{\small{Simulation of the quoted impacted price $\hat Y$ and of the unaffected price $\hat P$.}}
        \label{fig:evolutionPrixAvecSansImpact}
\end{figure}

\begin{figure}[h]
        \centering
        \includegraphics[height=5cm,width=\textwidth]{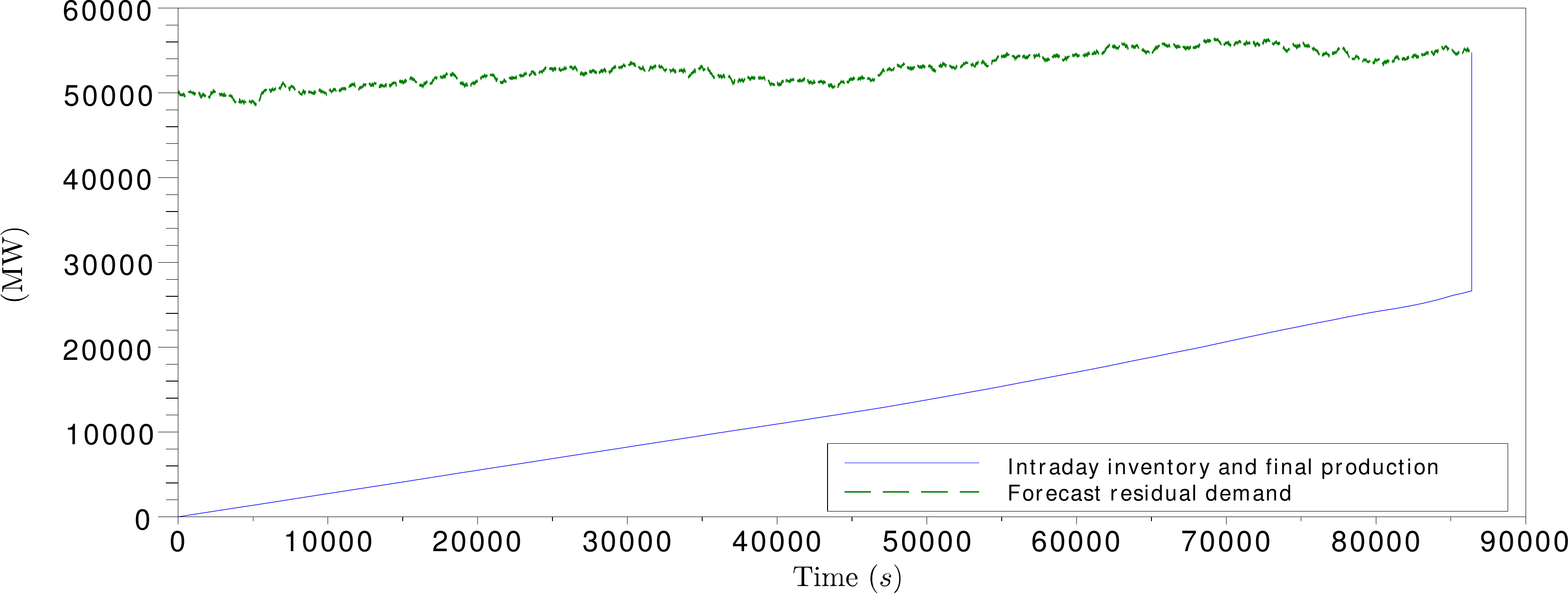}
        \caption{\small{Evolution of the inventory $\hat X$ and of the forecast residual demand $D$.}}
        \label{fig:evolutionXDPb}
\end{figure}

Figure \ref{fig:evolutionControlePb} represents the evolution of the trading rate control $(\hat q_t)_{t\in [0,T]}$ derived in \reff{optq} for a given trajectory of price and demand, and this is consistent with the martingale property as shown in Proposition~\ref{propmartinq}.    
Figure \ref{fig:evolutionPrixAvecSansImpact} represents a simulation of the quoted price $\hat Y_t$ with impact and of  the unaffected price $\hat P_t$.  Due to the buying strategy, i.e. positive $\hat q$, we observe that  the quoted price $\hat Y$ is larger than $\hat P$.  In Figure \ref{fig:evolutionXDPb}, we plot the evolution of the optimal inventory $(\hat X_t)_{t\in [0,T)}$, and of the forecast residual demand $(D_t)_{t\in [0,T]}$. We see that $\hat X_t$ is increasing, with a growth rate which looks constant as pointed out in Remark \ref{remmartinq}. At final time, if $\hat X_T$ $<$ $D_T$ (which is the case in our simulation), the agent uses her production leverage $\hat\xi_T$, and achieves a final inventory: $\hat X_T + \hat\xi_T$, which is represented by the peak at time $T$. 
From the expression \reff{optxiaux} of $\hat\xi_T$, the final  imbalance cost is  equal to
\beqs
D_T - \hat X_T - \hat\xi_T &=& \frac{\beta}{\eta+\beta} (D_T - \hat X_T),
\enqs
and is then   positive, as shown in Figure  \ref{fig:evolutionXDPb}.

 \vspace{1cm}

\clearpage

\section{Jumps in the residual demand forecast} \label{secjump}

\setcounter{equation}{0}
\setcounter{Assumption}{0} \setcounter{Theorem}{0}
\setcounter{Proposition}{0} \setcounter{Corollary}{0}
\setcounter{Lemma}{0} \setcounter{Definition}{0}
\setcounter{Remark}{0}

In this section, we incorporate the case where the residual demand forecast is subject to sudden changes induced by 
prediction errors on renewable production, which may be quite large.  
Our aim is to study the impact on the strategies obtained in the previous section, and we shall also neglect the delay in thermal plants production.

The sudden changes in the demand forecast are modeled via  a compound Poisson process $N_t$ $=$ $(N_t^+,N_t^-)_{t\geq 0}$ with intensity $\lambda$ $>$ $0$,  where $N_t^+$ is the counting process associated to positive jumps of the demand forecast with size $\delta^+$ $>$ $0$, occurring with probability $p^+$ $\in$ $[0,1]$, while $N_t^-$ is the counting process associated to negative  jumps of the demand forecast with size $\delta^-$ $<$ $0$, occurring with probability $p^-$ $=$ $1-p^+$.  We denote by $\delta$ $:=$ $\delta^+ p^+$ $+$ $\delta^- p^-$ the mean of the jump size of the demand forecast. 
The dynamics of the residual demand forecast $D$ is then  given by:
\beq \label{dynDjump}
dD_t &=& \mu dt + \sigma_d dB_t + \delta^+ dN_t^+ +  \delta^- dN_t^-,
\enq
where we add a jump component with respect to the model in \reff{dynD}.  Moreover, as soon as a jump in the residual demand forecast occurs, this is impacted  into the intraday electricity price since the  main producers are assumed to have access to the whole updated forecast.  We thus model the unaffected electricity price by:
\beq \label{prixjump}
\hat P_t &=& \hat P_0 + \sigma_0 W_t +  \pi^+ N_t^+ +  \pi^- N_t^-,
\enq
where we add with respect to the Bachelier model in \reff{Bachelier} a jump component of size $\pi^+$ $>$  $0$ (resp. $\pi^-$ $<$ $0$) 
when the jump on residual demand is positive (resp. negative), which means that a higher (resp. lower) demand induces an increase (resp. drop) of price.  
We denote by  $\pi$ $:=$ $\pi^+ p^+$ $+$ $\pi^- p^-$ the mean of the jump size of the intraday price.  
Given a trading rate $q$ $\in$ $\Ac$,  the dynamics of the quoted price $Y$ is then governed by
\beq \label{dynYjump}
dY_t &=& \nu q_t dt + \sigma_0 dW_t + \pi^+ dN_t^+  + \pi^- dN_t^-. 
\enq

By considering this simplified modeling of demand forecast subject to sudden shift  in terms of a Poisson process,  we  do not  have additional state variables with respect to the no jump case of the previous section.   Let us  then denote  by  $v$ $=$ $v^{(\lambda)}(t,x,y,d)$ the value function to the optimal execution problem \reff{defoptim} with cost functional $J$ $=$ $J^{(\lambda)}(t,x,y,d,q,\xi)$, 
where we stress the dependence in $\lambda$  for  taking into account jumps in demand forecast. 
The value function in the no jump case derived in the previous section is denoted by $v$ $=$ $v^{(0)}$.

As in the case with no jumps, there is no explicit solution to $v^{(\lambda)}$ due to the non negativity constraint on the final production: we shall first study  the auxiliary execution problem without sign constraint on the final production, then provide an approximate solution to the original one with an estimation of the induced  error approximation, and with some numerical illustrations. We  compare the results with the no jump  case by focusing on the impact of the jump components.

\subsection{Auxiliary optimal execution problem}

Similarly as in Subsection~\ref{secaux}, we consider the optimal execution problem without non negativity constraint on the final production, denoted by $\tilde v$ $=$ $\tilde v^{(\lambda)}(t,x,y,d)$. 

As in Theorem~\ref{thm1} for the case of the value function $\tilde v^{(0)}$ without jumps, we have an explicit solution to this auxiliary problem.

\begin{Theorem} \label{thm2}
The value function to the auxiliary optimization problem  is explicitly given by:
\beqs
& & \tilde v^{(\lambda)}(t,x,y,d)\\
&=& \tilde v^{(0)}(t,x,y,d) \\
& & +  \frac{\lambda}{2} \frac{r(\eta,\beta) (T-t) \big( \pi (T-t) + 2 \delta (\nu(T-t) + 2\gamma) \big)}{\big(r(\eta,\beta) + \nu\big)(T-t) + 2 \gamma} (d-x) \\
& & - \frac{\lambda}{2} \frac{(T-t)^2\big( \pi  - 2 r(\eta,\beta) \delta \big)}{\big(r(\eta,\beta) + \nu\big)(T-t) + 2 \gamma} y \\
& & +  \lambda\gamma \frac{p^+(\pi^+ -  r(\eta,\beta)\delta^+)^2+p^-(\pi^- -  r(\eta,\beta)\delta^-)^2} {\big(r(\eta,\beta) + \nu\big)^2} \ln\Big( 1 + \frac{(r(\eta,\beta) +  \nu)(T-t)}{2\gamma}\Big) \\
& & - \frac{\lambda}{2} \frac{p^+((\pi^+)^2- r(\eta,\beta) \delta^+(2\pi^++\nu\delta^+))+p^-((\pi^-)^2- r(\eta,\beta) \delta^-(2\pi^-+\nu\delta^-))}{r(\eta,\beta) + \nu} (T-t) \\
& & + \frac{\lambda r(\eta,\beta)}{2}\frac{2\nu\mu\delta  +\lambda((p^+)^2\delta^+(\pi^++\nu\delta^+)+(p^-)^2\delta^-(\pi^-+\nu\delta^-))}{ r(\eta,\beta) + \nu}(T-t)^2 \\
& & + \lambda^2 \gamma r(\eta,\beta) \frac{r(\eta,\beta)\delta^2 +2\nu p^+p^-\delta^+\delta^- -((p^+)^2\delta^+\pi^++(p^-)^2\delta^-\pi^-)}{(r(\eta,\beta)+ \nu)\big((r(\eta,\beta)+\nu)(T-t) + 2 \gamma\big)}(T-t)^2 \\
& & + \frac{2\lambda \gamma r^2(\eta,\beta) \mu\delta}{(r(\eta,\beta)+ \nu)\big((r(\eta,\beta)+\nu)(T-t) + 2 \gamma\big)}(T-t)^2 - \frac{\lambda^2\pi^2}{48\gamma}(T-t)^3 \\
& & + \frac{\lambda^2p^+p^-r(\eta,\beta)}{2}\frac{2\nu\delta^+\delta^-+\delta^-\pi^++\delta^+\pi^-}{(r(\eta,\beta)+\nu)(T-t) + 2\gamma}(T-t)^3\\
& & + \frac{1}{8}\frac{4r(\eta,\beta) \mu\lambda\pi - \lambda^2\pi^2}{(r(\eta,\beta)+\nu)(T-t) + 2\gamma}(T-t)^3,
\enqs
for $(t,x,y,d)$ $\in$ $[0,T]\times\R\times\R\times\R$, with an optimal trading rate given in feedback form by:
\beq 
\hat q_s^{(\lambda)}  & =& \hat q^{(\lambda)} (T-s,D_s^{t,d} - \hat X_s^{t,x,y,d},\hat Y_s^{t,x,y,d}), \;\;\; t \leq s \leq T  \nonumber \\
\hat q^{(\lambda)}(t,d,y) &:=& \hat q^{(0)}(t,d,y)  +  \lambda \frac{r(\eta,\beta)\delta t + \frac{\pi}{4\gamma}(r(\eta,\beta)+ \nu)t^2}{(r(\eta,\beta)+\nu)t + 2\gamma}  \nonumber \\
&=& \hat q^{(0)}(t,d+\lambda \delta  t,y+ \frac{\lambda}{2}\pi  t) + \frac{\lambda\pi}{4\gamma}t,  \label{hatqF} 
\enq
where $\hat q^{(0)}$ is the optimal trading rate given in \reff{optq} in the case with no jump in the demand forecast.   
Here $(\hat X^{t,x,y,d},\hat Y^{t,x,y,d},D^{t,d})$ denotes the solution to \reff{dynX}-\reff{dynYjump}-\reff{dynDjump} when using the feedback control $\hat q^{(\lambda)}$, 
and starting from  $(x,y,d)$ at time $t$.  Finally, the optimal production quantity is given by:
\beq \label{hatxiF}
\hat \xi_T^{(\lambda)} &=&  \frac{\eta}{\eta + \beta} \big(D_T^{t,d} - \hat X_T^{t,x,y,d}\big). 
\enq
\end{Theorem} 
{\bf Proof.} See Appendix.
\ep
 
\vspace{3mm}

\noindent {\bf Interpretation.}   The expression of the optimal trading rate $\hat q_s^{(\lambda)}$,  $s$ $\in$ $[t,T]$, as 
\beqs
\hat q_s^{(\lambda)} &=& \hat q_s^{(0)} +   
\lambda \frac{r(\eta,\beta)\delta (T-s) + \frac{\pi}{4\gamma}(r(\eta,\beta)+ \nu)(T-s)^2}{(r(\eta,\beta)+\nu)(T-s)+ 2\gamma}, 
\enqs
where $\hat q_s^{(0)}$ $=$ $\hat q^{(0)}(T-s,D_s^{t,d}-\hat X_s^{t,x,y,d},\hat Y_s^{t,x,y,d})$ represents the optimal trading rate that the agent would use if she believes that the demand forecast will not jump, shows that under the information knowledge  about jumps, the agent will purchase more (resp. less) electricity shares and this impact is all the larger, the larger the intensity $\lambda$ of jumps, and the positive (resp. negative) mean $\delta$ and $\pi$ of jump size in demand forecast and price are.  On the other hand,  the expression of $\hat q_s^{(\lambda)}$  as the sum of two terms:
\beq \label{linearqF}
\hat q_s^{(\lambda)} &=& \hat q^{(0)}\big(T-s,D_s^{t,d}+ \lambda\delta(T-s),\hat Y_s^{t,x,y,d} + \frac{\lambda}{2} \pi (T-s) \big) \; 
+ \;  \frac{\lambda\pi}{4\gamma}(T-s),
\enq
can be interpreted as follows. The first term is analog to the optimal trading rate in the no jump case, with an adjustment 
$\lambda\delta(T-s)$ in the demand, which represents the expectation of the demand jump size up to  the final horizon,  
and an adjustment $\frac{\lambda}{2} \pi (T-s)$ on the price, which represents half of the expectation of the price jump size up to the final horizon.  
The second term,  $\frac{\lambda\pi}{4\gamma}(T-s)$,  is deterministic, and linear in time, and we shall see on the simulations  for some parameter values  that it can be dominant with respect to  the first stochastic term.   Moreover, notice that the equilibrium relation \reff{eq:qnojump2}  in the no jump case between forecast intraday price and forecast marginal cost of production does not hold anymore in the presence of jumps, except at terminal date $T$:
\beq \label{relequil2}
\hat Y_T^{t,x,y,d} +  2\gamma \hat q_T^{(\lambda)} &=&  c'(\hat \xi_T^{(\lambda)}). 
\enq
\ep

\vspace{2mm}

The unaffected price $\hat P$ in \reff{prixjump}  is no more a martingale in presence of jumps, except when $\pi$ $=$ $0$.  It is actually a supermartingale  when $\pi$ $<$ $0$ (predominant negative jumps), and submartingale when $\pi$ $>$ $0$ (predominant positive jumps).  
The next result shows that the optimal trading rate inherits the converse submartingale or supermartingale property of the price process.

\begin{Proposition} \label{propsupermartinq}
The optimal trading rate process $(\hat q_s^{(\lambda)})_{t\leq s \leq T}$ in \reff{hatqF} is a supermartingale if $\pi$ $>$ $0$, 
and a submartingale if $\pi$ $<$ $0$. More precisely, the process 
$\{\hat q_s^{(\lambda)} + \frac{\lambda\pi}{2\gamma}(s-t), t\leq s\leq T\}$ is a martingale. 
\end{Proposition}
{\bf Proof.} Notice that $N_t^\pm$ is a Poisson process with intensity $\lambda p^\pm$, and let us introduce the compensated martingale Poisson process 
$\tilde N_t^\pm$ $=$ $N_t - \lambda p^\pm t$. By applying It\^o's formula to the trading rate process $\hat q_s^{(\lambda)}$ $=$ 
$\hat q^{(\lambda)}(T-s,D_s^{t,d}-\hat X_s^{t,x,y,d},\hat Y_s^{t,x,y,d})$, $t\leq s\leq T$, and from the dynamics \reff{dynX}, \reff{dynDjump} and \reff{dynYjump}, we have: 
\beqs
d\hat q_s^{(\lambda)} &=&  \Big[ - \Dt{\hat q^{(\lambda)}} + (\mu - \hat q^{(\lambda)})\Dd{\hat q^{(\lambda)}} + \nu \hat q^{(\lambda)} 
\Dy{\hat q^{(\lambda)}} \nonumber \\
& & \;\;\;\;\;  + \; \lambda p^+\big( \hat q^{(\lambda)}(.,.+\delta^+, . + \pi^+) - \hat q^{(\lambda)}\big) \nonumber  \\
& & \;\;\;\;\;  + \; \lambda p^-\big( \hat q^{(\lambda)}(.,.+\delta^-, . + \pi^-) - \hat q^{(\lambda)}\big) \Big] (T-s,D_s^{t,d}-\hat X_s^{t,x,y,d},\hat Y_s^{t,x,y,d}) ds \nonumber  \\
& &  \; + \; \Dd{\hat q^{(\lambda)}}(T-s,D_s^{t,d}-\hat X_s^{t,x,y,d},\hat Y_s^{t,x,y,d})  \sigma_d dB_s \nonumber \\
& &  \; + \;   \Dy{\hat q^{(\lambda)}}(T-s,D_s^{t,d}-\hat X_s^{t,x,y,d},\hat Y_s^{t,x,y,d}) \sigma_0 dW_s \nonumber \\
& &  \; + \; \big[ \hat q^{(\lambda)}(T-s,D_{s^-}^{t,d} + \delta^+ - \hat X_s^{t,x,y,d},\hat Y_{s^-}^{t,x,y,d} + \pi^+) \nonumber \\
& & \hspace{3cm}  - \;   \hat q^{(\lambda)}(T-s,D_{s^-}^{t,d}  - \hat X_s^{t,x,y,d},\hat Y_{s^-}^{t,x,y,d}) \big] d \tilde N_s^+ \nonumber \\
& &  \; + \; \big[ \hat q^{(\lambda)}(T-s,D_{s^-}^{t,d} + \delta^- - \hat X_s^{t,x,y,d},\hat Y_{s^-}^{t,x,y,d} + \pi^-) \nonumber \\
& & \hspace{3cm}  - \;   \hat q^{(\lambda)}(T-s,D_{s^-}^{t,d}  - \hat X_s^{t,x,y,d},\hat Y_{s^-}^{t,x,y,d}) \big] d \tilde N_s^-. \nonumber
\enqs 
Now,  from the expression \reff{hatqF} of $\hat q^{(\lambda)}(t,d,y)$, we see that:
\beqs
- \Dt{\hat q^{(\lambda)}} + (\mu - \hat q^{(\lambda)})\Dd{\hat q^{(\lambda)}} + \nu \hat q^{(\lambda)} \Dy{\hat q^{(\lambda)}}&& \\ 
+\lambda\big( p^+\hat q^{(\lambda)}(.,.+\delta^+, . + \pi^+) +  p^-\hat q^{(\lambda)}(.,.+\delta^-, . + \pi^-)  - \hat q^{(\lambda)}\big) 
&=& -  \frac{\lambda\pi}{2\gamma},
\enqs
and then: 
\beq
d\hat q_s^{(\lambda)} &=&   -  \frac{\lambda\pi}{2\gamma}  ds  \nonumber \\
& &  \; + \;  \frac{r(\eta,\beta)\sigma_d }{(r(\eta,\beta) + \nu)(T-s) + 2 \gamma} dB_s 
- \frac{\sigma_0 }{(r(\eta,\beta) + \nu)(T-s) + 2 \gamma} dW_s \nonumber \\
& &   \; + \;  \frac{r(\eta,\beta)\delta^+ - \pi^+}{(r(\eta,\beta) + \nu)(T-s) + 2 \gamma}  d \tilde N_s^+  
+ \frac{r(\eta,\beta)\delta^- - \pi^-}{(r(\eta,\beta) + \nu)(T-s) + 2 \gamma}  d \tilde N_s^-. \label{dynqF}
\enq
This proves  the  required assertions of the proposition. 
\ep

\begin{Remark} \label{remsupermartinq}
{\rm The above supermartingale (or submartingale) property  implies in particular that the mean of the  optimal trading rate process 
$(\hat q_s^{(\lambda)})_{0\leq s\leq T}$ is decreasing (or increasing) in time, and so that the trajectory of the optimal inventory mean $\E[\hat X_s^{0,x,y,d}]$, 
$0\leq s\leq T$, is concave (or convex).  Moreover, from the martingale property of $\hat q_s^{(\lambda)}+\frac{\lambda\pi}{2\gamma}s$, $0\leq s\leq T$, we have:  
$\E[\hat q_s^{(\lambda)}]$ $=$  $\hat q^{(\lambda)}(T,d-x,y)$ $-$ $\frac{\lambda\pi}{2\gamma}s$  
for $0\leq s\leq T$.  Fix $d,x,y$, and let us then denote by  $\bar s^{(\lambda)}$ $:=$ $\frac{2\gamma}{\lambda\pi}\hat q^{(\lambda)}(T,d-x,y)$, which is explicitly written as:
\[
\bar s^{(\lambda)} = \frac{T}{2} + \frac{1}{\lambda\pi} \frac{\big(r(\eta,\beta)\mu + \lambda(r(\eta,\beta)\delta - \frac{\pi}{2}) \big)T + r(\eta,\beta) (d-x) - y}{1 + \frac{(r(\eta,\beta)+\nu)T}{2\gamma}}
\]
We have the following cases:
\begin{itemize}
\item $\bar s^{(\lambda)}$ $\leq$ $0$ and $\pi$ $>$ $0$: this  may arise for large $y$, or $d$ $<<$ $x$, or $r(\eta,\beta)\delta << \pi/2$. In this extreme case, $\frac{d \E[\hat X_s^{0,x,y,d}]}{ds}$ $=$  $\E[\hat q_s^{(\lambda)}]$ $\leq$ $0$ for $0\leq s\leq T$, i.e. the trajectory of $\E[\hat X_s^{0,x,y,d}]$, $0\leq s \leq T$, is decreasing, which means that the agent will ``always''  sell electricity shares since she takes advantage of high price, in order to decrease her inventory for approaching the demand, and because in average,  the jump size of the demand is much lower than the positive jump size of the price. 
\item $\bar s^{(\lambda)}$ $\leq$ $0$ and $\pi$ $<$ $0$: this may arise for  small $y$, or $d$ $>>$ $x$, or $r(\eta,\beta)\delta >> \pi/2$. In this extreme case, $\frac{d \E[\hat X_s^{0,x,y,d}]}{ds}$ $=$  $\E[\hat q_s^{(\lambda)}]$ $\geq$ $0$ for $0\leq s\leq T$, i.e. the trajectory of $\E[\hat X_s^{0,x,y,d}]$, $0\leq s \leq T$, is increasing, which means that  the agent will ``always'' buy electricity shares 
since she takes advantage of low price, in order to increase her inventory for approaching the demand, and because in average,  the jump size of the price is much lower than the jump size of the demand.  
\item $\bar s^{(\lambda)}$ $\geq$ $T$ and $\pi$ $>$ $0$: this may arise for $r(\eta,\beta)\delta$ $>>$ $\pi/2$, $d>>x$ or small $y$. In this other extreme case,  the trajectory of $\E[\hat X_s^{0,x,y,d}]$, $0\leq s \leq T$, is increasing,  which means that  the agent will ``always'' buy  electricity shares at low price  in order to approach the residual demand at final time. 
\item $\bar s^{(\lambda)}$ $\geq$ $T$ and $\pi$ $<$ $0$: this may arise for $r(\eta,\beta)\delta$ $<<$ $\pi/2$, $d<<x$ or large $y$. 
The trajectory of $\E[\hat X_s^{0,x,y,d}]$, $0\leq s \leq T$, is decreasing, which means that the agent will ``always'' sell  electricity shares at high price  in order to approach the residual demand at final time. 
\item $0$ $<$ $\bar s^{(\lambda)}$ $<$ $T$: in this regular case, it is interesting to comment on the two subcases:
\begin{itemize}
\item if $\pi$ $>$ $0$, the trajectory of $s$ $\mapsto$ $\E[\hat X_s^{0,x,y,d}]$ is increasing for $s$ $\leq$ $\bar s^{(\lambda)}$ and then decreasing for $\bar s^{(\lambda)}$ $<$ $s$ $\leq$ $T$.  This means that the agent starts by purchasing  electricity shares for taking profit of the positive price jumps (which have more impact than the negative price jumps as $p^+\pi^++p^-\pi^->0$), and then resells shares in order to achieve the equilibrium relation \reff{relequil2}. 
\item if $\pi$ $<$ $0$, i.e. the negative jumps have more impact than the positive ones: the agent starts by selling electricity shares and then purchases shares. 
\end{itemize} 
\end{itemize}
\ep
}
\end{Remark}

\subsection{Approximate solution}

We turn back to the original optimal execution problem with the non negativity constraint on the final production, and as in Section~\ref{secapprox}, we use the approximate strategy consisting in 
the trading rate $\hat q^{(\lambda)}$ derived in \reff{hatqF}, and of the truncated nonnegative final production:
\beqs
\tilde\xi_T^{(\lambda),*} & := & \hat\xi_T^{(\lambda)}  \ind_{_{\hat\xi_T^{(\lambda)} \geq 0}} \; = \; \hat\xi^+(D_T^{t,d} - \hat X_T^{t,x,y,d}), 
\enqs
with $\hat\xi_T^{(\lambda)}$ given in \reff{hatxiF}.   We measure the relevance of this strategy  $(\hat q^{(\lambda)},\tilde\xi_T^{(\lambda),*})$ $\in$ $\Ac\times L_+^0(\Fc_T)$  by estimating the  induced error:
 \beqs
\Ec_1^{(\lambda)}(t,x,y,d) & := &  J^{(\lambda)}(t,x,y,d;\hat q^{(\lambda)},\tilde\xi_T^{(\lambda),*}) - v^{(\lambda)}(t,x,y,d),
 \enqs
for $(t,x,y,d)$ $\in$ $[0,T]\times\R\times\R\times\R$, and also measure the approximation error on the value functions:
\beqs
\Ec_2^{(\lambda)}(t,x,y,d) & := & v^{(\lambda)}(t,x,y,d) - \tilde v^{(\lambda)}(t,x,y,d). 
\enqs

\begin{Proposition}
For all $(t,x,y,d)$ $\in$ $[0,T]\times\R\times\R\times\R$, we have
\beq 
0 \; \leq \; \Ec_i^{(\lambda)}(t,x,y,d) & \leq & \frac{\eta r(\eta,\beta)}{2\beta} V(T-t) 
\E \Big[ \psi \Big( \frac{m^{(\lambda)}(T-t,d-x,y) + \Sigma_T^{-,t}}{\sqrt{V(T-t)}}\Big) \Big] \label{bornsaut} 
\enq
for $i$ $=$ $1,2$, where $\psi$,  $m$, $V$ are defined  in Proposition \ref{erreurnopanne}, 
\beq
m^{(\lambda)}(t,d,y) &=& m\Big(t,d,y + \lambda\big(\frac{\pi}{2} - r(\eta,\beta)\delta\big)t \Big)  \label{defmF} \\
& & \;+ \;  \lambda \frac{ r(\eta,\beta)\delta - \pi}{r(\eta,\beta)+\nu} \Big[ t   - \frac{2\gamma}{r(\eta,\beta)+\nu} \ln\Big( 1 + \frac{r(\eta,\beta) +  \nu}{2\gamma}t\Big) \Big],
\nonumber 
\enq
and 
\beqs
\Sigma_T^{-,t} &=&   \int_t^T \frac{\delta^- ( \nu(T-s)+2\gamma) + \pi^-(T-s) } {(r(\eta,\beta) + \nu)(T-s) + 2 \gamma} dN_s^- \;\; \leq \; 0, \;\; a.s. 
\enqs
\end{Proposition}
{\bf Proof.} By the same arguments as in Proposition~\ref{erreurnopanne}, we have
\beqs
0 \leq \;  \Ec_i^{(\lambda)}(t,x,y,d)  & \leq & \Ec^{(\lambda)}(t,x,y,d) \; := \; J^{(\lambda)}(t,x,y,d;\hat q^{(\lambda)},\tilde\xi_T^{(\lambda),*}) - J^{(\lambda)}(t,x,y,d;\hat q^{(\lambda)},\hat\xi_T^{(\lambda)}),
\enqs
for $i$ $=$ $1,2$, and 
\beq
\Ec^{(\lambda)}(t,x,y,d)   &=& \frac{\eta r(\eta,\beta)}{2\beta} \E \Big[  \big(D_T^{t,d} - \hat X_T^{t,x,y,d} \big)^2 \ind_{D_T^{t,d} - \hat X_T^{t,x,y,d} < 0} \Big],  \label{expressinterEcjump}
\enq
for $(t,x,y,d)$ $\in$ $[0,T]\times\R\times\R\times\R$. Now, recall from \reff{dynqF} that: 
\beqs
d\hat q_s^{(\lambda)} &=& - \lambda \Big[  \frac{\pi}{2\gamma} + \frac{r(\eta,\beta)\delta - \pi}{(r(\eta,\beta)+\nu)(T-s) + 2 \gamma}  \Big] ds  \\
 & & \; + \; \frac{r(\eta,\beta)\sigma_d }{(r(\eta,\beta) + \nu)(T-s) + 2 \gamma} dB_s   \; - \;    \frac{\sigma_0 }{(r(\eta,\beta) + \nu)(T-s) + 2 \gamma} dW_s  \\
 & & \; + \; \frac{r(\eta,\beta)\delta^+ - \pi^+}{(r(\eta,\beta) + \nu)(T-s) + 2 \gamma} dN_s^+ + \; \frac{r(\eta,\beta)\delta^- - \pi^-}{(r(\eta,\beta) + \nu)(T-s) + 2 \gamma} dN_s^-,
\enqs
where we write the dynamics directly  in terms of the Poisson processes $N^\pm$.  By integration, 
we deduce the (path-dependent) expression of $\hat q_s^{(\lambda)}$, $t\leq s \leq T$: 
\beqs
\hat q_s^{(\lambda)} &=& \hat q_t^{(\lambda)}   -  \frac{\lambda\pi}{2\gamma} (s-t)  +
 \frac{\lambda \big(r(\eta,\beta)\delta - \pi\big)}{r(\eta,\beta) + \nu} \ln \Big( \frac{(r(\eta,\beta)+\nu)(T-s) + 2 \gamma}{(r(\eta,\beta)+\nu)(T-t) + 2 \gamma}\Big) \\
 & & \; + \; \int_t^s \frac{r(\eta,\beta)\sigma_d }{(r(\eta,\beta) + \nu)(T-u) + 2 \gamma} dB_u 
 - \int_t^s   \frac{\sigma_0 }{(r(\eta,\beta) + \nu)(T-u) + 2 \gamma} dW_u \\
 & & \; + \; \int_t^s \frac{r(\eta,\beta)\delta^+ - \pi^+}{(r(\eta,\beta) + \nu)(T-u) + 2 \gamma} dN_u^+ +  \; \int_t^s \frac{r(\eta,\beta)\delta^- - \pi^-}{(r(\eta,\beta) + \nu)(T-u) + 2 \gamma} dN_u^-,
\enqs
with  $\hat q_t^{(\lambda)}$ $=$ $\hat q^{(\lambda)}(T-t,d-x,y)$. 
We thus  obtain the expression of the final spread between demand and inventory: 
\beq 
D_T^{t,d} - \hat X_T^{t,x,y,d} &=& d- x +  \mu(T-t) +  \int_t^T  \sigma_d dB_s  + \int_t^T \delta^+ dN_s^+ +   \int_t^T \delta^- dN_s^-  -   \int_t^T \hat q_s^{(\lambda)} ds  \nonumber \\
&=& m^{(\lambda)}(T-t,d-x,y)   \nonumber \\
& &  + \;  \int_t^T \frac{\sigma_d(\nu(T-s)+2\gamma)}{(r(\eta,\beta) + \nu)(T-s) + 2 \gamma} dB_s  \; 
+ \; \int_t^T \frac{\sigma_0 (T-s)}{(r(\eta,\beta) + \nu)(T-s) + 2 \gamma} dW_s \nonumber \\
& & \; + \; \int_t^T \frac{\delta^+ ( \nu(T-s)+2\gamma) + \pi^+(T-s) } {(r(\eta,\beta) + \nu)(T-s) + 2 \gamma} dN_s^+  \nonumber \\
& &  \; + \; \int_t^T \frac{\delta^- ( \nu(T-s)+2\gamma) + \pi^-(T-s) } {(r(\eta,\beta) + \nu)(T-s) + 2 \gamma} dN_s^-,  
\label{spreadjump}
\enq
by Fubini's theorem, and where
\beqs
m^{(\lambda)}(t,d,y) &:=& d + \mu t - t \hat q^{(\lambda)}(t,d,y) + \frac{\lambda\pi}{2\gamma} \int_0^t s  ds  \\
& &  \;\; - \;  \frac{\lambda \big(r(\eta,\beta)\delta - \pi\big)}{r(\eta,\beta) + \nu} \int_0^t 
\ln \Big( \frac{(r(\eta,\beta)+\nu) s + 2 \gamma}{(r(\eta,\beta)+\nu) t + 2 \gamma}\Big) ds, 
\enqs
is explicitly written as in \reff{defmF} after some straightforward calculation. 
Denoting by $\Delta_T^{t,x,y,d}$ the continuous part of $D_T^{t,d} - \hat X_T^{t,x,y,d}$ consisting in the three first terms in the rhs of \reff{spreadjump}, and by $\Sigma_T^{+,t}$, $\Sigma_T^{-,t}$  the jump parts consisting in the two  last terms of \reff{spreadjump}, so that 
\beqs
D_T^{t,d} - \hat X_T^{t,x,y,d} &=&  \Delta_T^{t,x,y,d} + \Sigma_T^{+,t} + \Sigma_T^{-,t}, 
\enqs
we notice that $\Delta_T^{t,x,y,d}$ follows a normal distribution law with mean $m^{(\lambda)}(T-t,d-x,y)$ and variance $V(T-t)$, independent of  
$\Sigma_T^{\pm,t}$.   Then, conditionally on $\Sigma_T^{\pm,t}$, $D_T^{t,d} - \hat X_T^{t,x,y,d}$  follows a 
normal distribution law  with mean $m^{(\lambda)}(T-t,d-x,y)$ $+$ $\Sigma_T^{+,t}$ $+$ $\Sigma_T^{-,t}$, and variance $V(T-t)$, and this implies from \reff{expressinterEcjump} that: 
\beqs
\Ec^{(\lambda)}(t,x,y,d)   &=& \frac{\eta r(\eta,\beta)}{2\beta} V(T-t)  \E \Big[ \psi \Big( \frac{m^{(\lambda)}(T-t,d-x,y) + \Sigma_T^{+,t} + \Sigma_T^{-,t}}{\sqrt{V(T-t)}}\Big) \Big]    \\ 
& \leq &   \frac{\eta r(\eta,\beta)}{2\beta}  V(T-t) \E \Big[  \psi \Big( \frac{m^{(\lambda)}(T-t,d-x,y) + \Sigma_T^{-,t} }{\sqrt{V(T-t)}}\Big) \Big], 
\enqs
since $\Sigma_T^{+,t}$ $\geq$ $0$ a.s. and $\psi$ is non-increasing. 
\ep

\vspace{3mm}

\noindent {\bf  Comments on the approximation error.}  Let us discuss about the accuracy of the upper bound in \reff{bornsaut}: 
\beqs
\bar \Ec^{(\lambda)}(T-t,d-x,y)   &:=& \frac{\eta r(\eta,\beta)}{2\beta}  V(T-t) \E \Big[  \psi \Big( \frac{m^{(\lambda)}(T-t,d-x,y) + \Sigma_T^{-,t} }{\sqrt{V(T-t)}}\Big) \Big], 
\enqs
First, notice that  $m^{(\lambda)}(T-t,d-x,y)$ $+$ $\Sigma_T^{-,t}$  $\sim$  $m(T-t,d-x,y)$ a.s.  in the limiting regimes where $T-t$ goes to zero, $d-x$ or $y$ goes to infinity. Therefore, by dominated 
convergence theorem, $\bar \Ec^{(\lambda)}(T-t,d-x,y)$ converges to zero in these limiting regimes as in the no jump case. However,  we are not able to  derive an  asymptotic limit  as in the no jump  case of Proposition~\ref{properrorlimite},   except when $\Sigma_T^{-,t}$ $=$ $0$, i.e.  $\delta^-$ $=$ $\pi^-$ $=$ $0$, fo which we get the same asymptotic limit. 
Actually,  in the presence of negative jumps on the demand, it is  intuitively clear that  our approximation should be less accurate than in the no jump case since the probability for  the residual demand  to stay above the final inventory  is decreasing. Anyway,  the explicit strategies  $(\hat q^{(\lambda)},\tilde\xi_T^{(\lambda),*})$ still provide a very accurate  approximation of the optimal strategies at least in these limiting regimes, as illustrated in the next paragraph.

\subsection{Numerical results}

We plot trajectories of some relevant quantities that we simulate with the same set of parameters as in Paragraph~\ref{simul}:  
$\sigma_0=1/60$ \euro{}$\cdot$(MW)$^{-1}\cdot s^{-1/2}$, $\sigma_d=1000/60$ MW$\cdot s^{-1/2}$, 
$\beta=0.002$ \euro{}$\cdot$(MW)$^{-2}$, $\eta=200$ \euro{}$\cdot$(MW)$^{-2}$,  $\mu=0$ MW$\cdot s^{-1}$,  $\rho=0.8$, 
$\nu=4.00\cdot 10^{-5}$\euro{}$\cdot$(MW)$^{-2}$, $\gamma=2.22$\euro{}$\cdot s\cdot$(MW)$^{-2}$,  $T$ $=$ $24$h, $X_0$ $=$ $0$, $D_0=50000$ MW and  $Y_0=50$ \euro{}$\cdot$(MW)$^{-1}$.  Moreover, we fix the probability of positive jumps, $p^+=1$ (then all jumps are positive: $p^-=0$), and the following values for the jump components: $\lambda=1.5/(3600\cdot24)$ $s^{-1}$, $\pi^+=10$ \euro{}$\cdot$(MW)$^{-1}$, $\delta^+=1500$ MW.

For such parameter values, we observe two occurrences of jumps on the trajectories of the demand of price.  Moreover, the probability $\P[\hat X_T > D_T]$ is bounded above by $2.92\times 10^{-16}$,  the error 
$\bar\Ec^{(\lambda)}(0,D_0 -X_0,Y_0)$ is bounded by $2.66\times 10^{-5}$\euro{}, and 
\beqs
\tilde v^{(\lambda)}(0,X_0,Y_0,D_0) &=& 2020950 \text{\euro{}.}
\enqs
The executed strategy $(\hat q^{(\lambda)},\hat\xi_T^{(\lambda),*})$ can then be considered as very close to the optimal strategy.  This has to be compared with the numerical result obtained in the previous section in the no jump case where we obtained a lower expected total cost: $\tilde v(0,X_0,Y_0,D_0)$ $=$ $1916700 \text{\euro{}.}$

\begin{figure}[h]
        \centering
        \includegraphics[height=4cm,width=\textwidth]{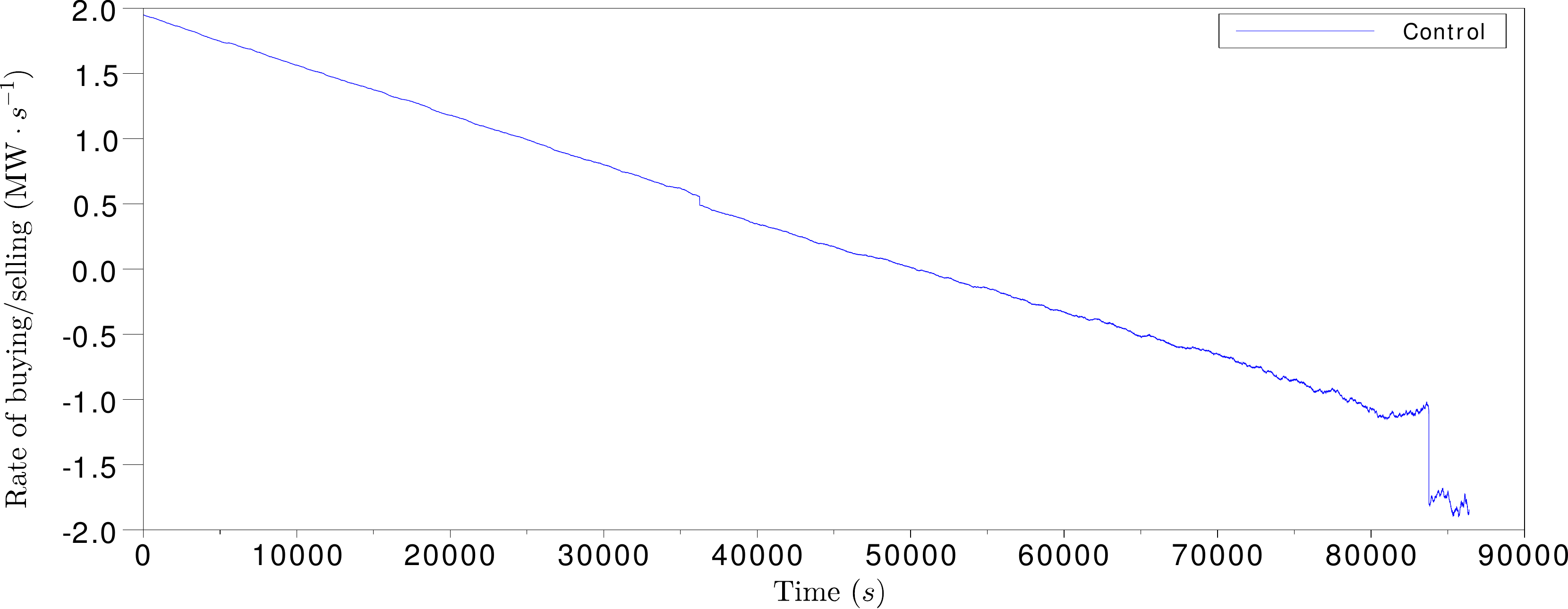}
        \caption{\small{Evolution of the trading rate control $\hat q^{(\lambda)}$}}
        \label{fig:evolutionControlePbSauts}
\end{figure}

\begin{figure}[h]
        \centering
        \includegraphics[height=4cm,width=\textwidth]{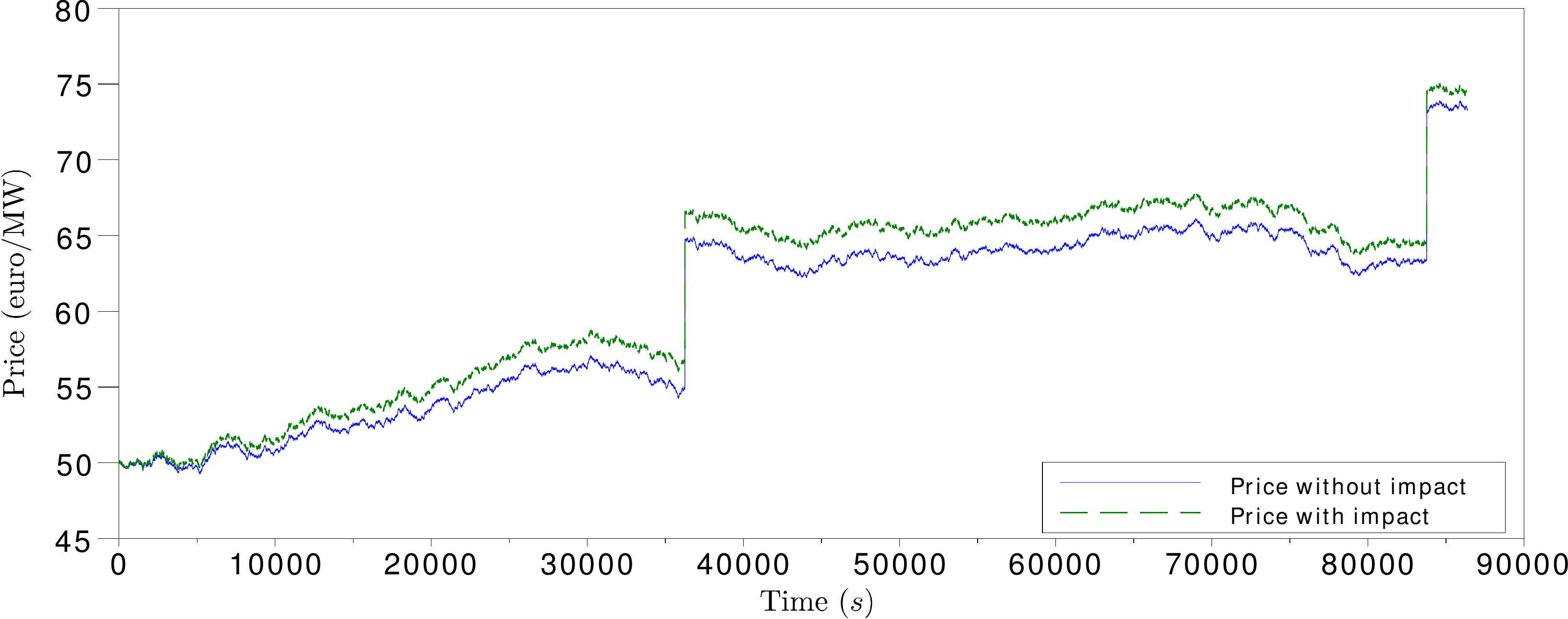}
        \caption{\small{Simulation of the quoted impacted price $\hat Y$ and of the unaffected price $\hat P$}}
        \label{fig:evolutionPrixAvecSansImpactSauts}
\end{figure}

\begin{figure}[h]
        \centering
        \includegraphics[height=4cm,width=\textwidth]{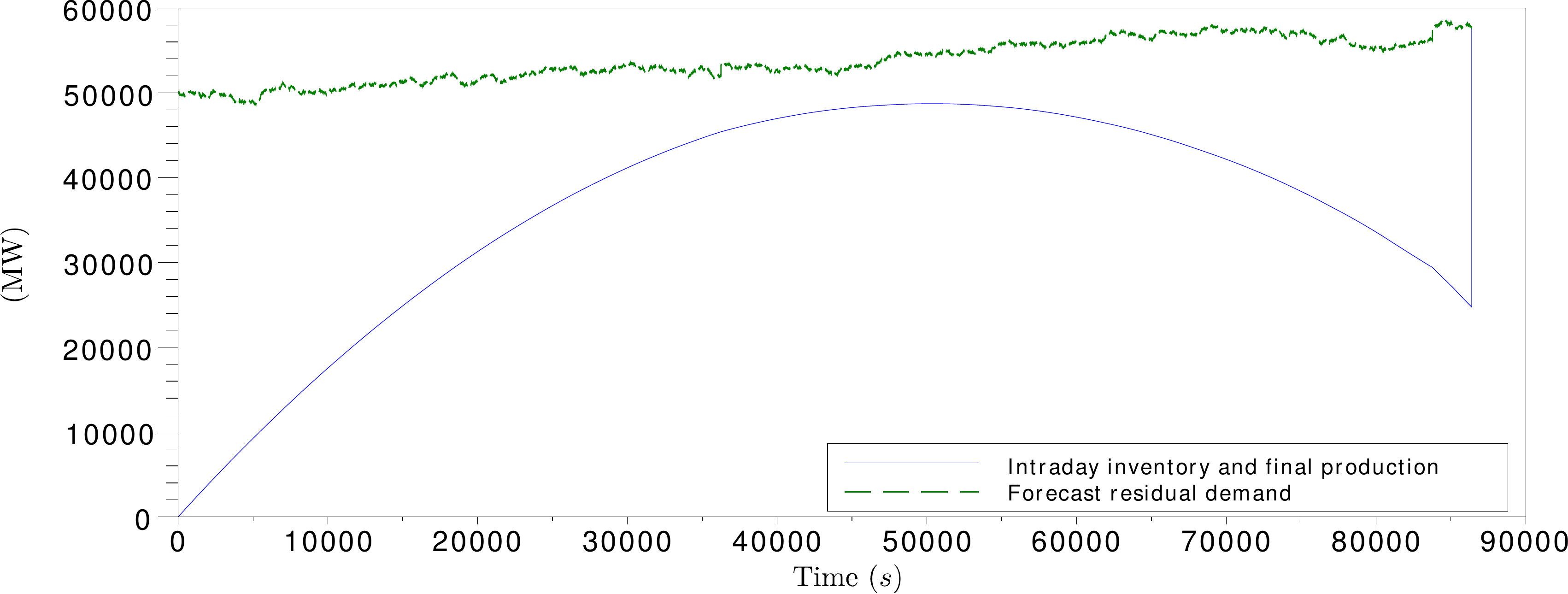}
        \caption{\small{Evolution of the inventory $\hat X$ and of the forecast residual demand $D$}}
        \label{fig:evolutionXDPbSauts}
\end{figure}

Figure \ref{fig:evolutionControlePbSauts} represents the evolution of the trading rate $(\hat q_t^{(\lambda)})_{t\in [0,T]}$, and we see that it is decreasing  consistently with the supermartingale property in Proposition~\ref{propsupermartinq}. Actually, we observe that the deterministic part in  \reff{linearqF}, which is linear in time, dominates the stochastic part. The interpretation of the strategy is the following: since  positive price jumps are expected, the agent purchases a large number of shares in electricity with the hope to sell it later at a higher price thanks to the possible occurrence of a positive jump.   
At the price jump times, which can be visualized in Figure \ref{fig:evolutionPrixAvecSansImpactSauts}, we notice that the control $\hat q^{(\lambda)}$ reacts by a decrease in the trading rate. The reaction to the second jump is more sensible than to the first jump since it occurs a short time before the final horizon $T$, where the objective is also to achieve the equilibrium relation \reff{relequil2} between price and  marginal cost.  Finally, we observe clearly in Figure~\ref{fig:evolutionXDPbSauts} the concavity of the trajectory of the optimal inventory process $(\hat X_t)_{t\in [0,T)}$, as expected from Remark \ref{remsupermartinq}. This emphasizes the double objective of the agent: on one hand, the purchase of  electricity shares for taking profit of the positive  price jumps, and on the other hand the resale of electricity shares for attaining the equilibrium relation between price and  marginal cost at terminal date.  We also plot the production $\hat\xi_T$ at the final time $T$ on Figure~\ref{fig:evolutionXDPbSauts},  and observe as in the no jump case that the imbalance cost $D_T - \hat X_T -\hat\xi_T$ is positive. 

\vspace{1cm}

\vspace{2cm}

\vspace{1cm}

\vspace{1cm}

Next, we plot trajectories with the same set of parameters, but with $p^+=0.3$ (i.e. $p^-=0.7$), $\pi^-=-10$ \euro{}$\cdot$(MW)$^{-1}$, $\delta^-=-1500$ MW. There are, in average, more negative than positive jumps. Now
\beqs
\tilde v^{(\lambda)}(0,X_0,Y_0,D_0) &=& 1756330 \text{\euro{}.}
\enqs

\begin{figure}[h]
        \centering
        \includegraphics[height=4cm,width=\textwidth]{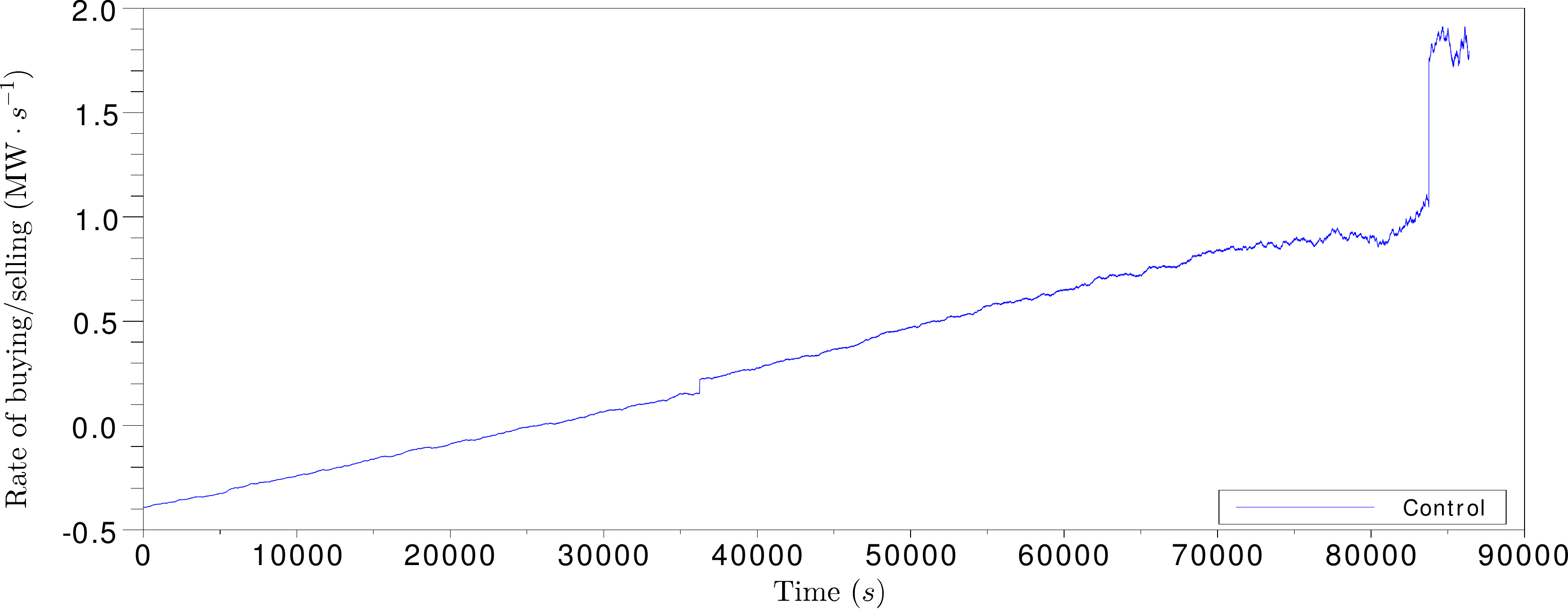}
        \caption{\small{Evolution of the trading rate control $\hat q^{(\lambda)}$}}
        \label{fig:evolutionControlePbSautsNegatifsDominants}
\end{figure}

\begin{figure}[h]
        \centering
        \includegraphics[height=4cm,width=\textwidth]{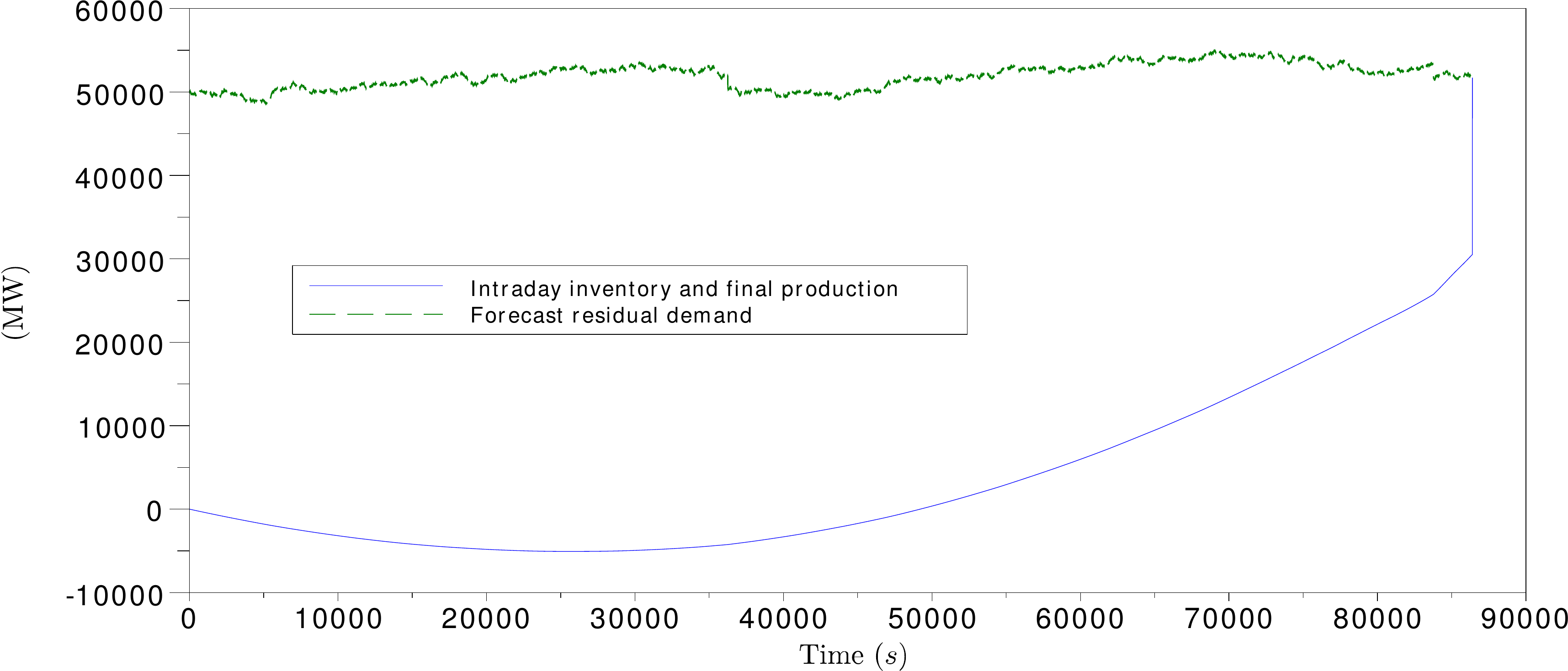}
        \caption{\small{Evolution of the inventory $\hat X$ and of the forecast residual demand $D$}}
        \label{fig:evolutionXDPbSautsNegatifsDominants}
\end{figure}

Figure \ref{fig:evolutionControlePbSautsNegatifsDominants} shows that the trading rate $(\hat q_t^{(\lambda)})_{t\in [0,T]}$ is increasing, which is consistent with the submartingale property in Proposition~\ref{propsupermartinq}: the deterministic part in \reff{linearqF} dominates the stochastic part. Since negative jumps are more expected than negative jumps are, the agent first sells a large number of shares in electricity with the hope to buy it later at a lower price thanks to the possible occurrence of jumps, that should be mainly negative. 
Here, the control reacts to the negative price jumps by an increase in the trading rate. Finally, in Figure \ref{fig:evolutionXDPbSautsNegatifsDominants} we observe the convexity of the trajectory of the optimal inventory $(\hat X_t)_{t\in [0,T)}$ process, as expected from Remark \ref{remsupermartinq}. We also plot the production $\hat\xi_T$ at the final time $T$ on that figure. 

\vspace{1cm}

\vspace{1cm}

\begin{figure}[h]
        \centering
        \includegraphics[height=4cm,width=\textwidth]{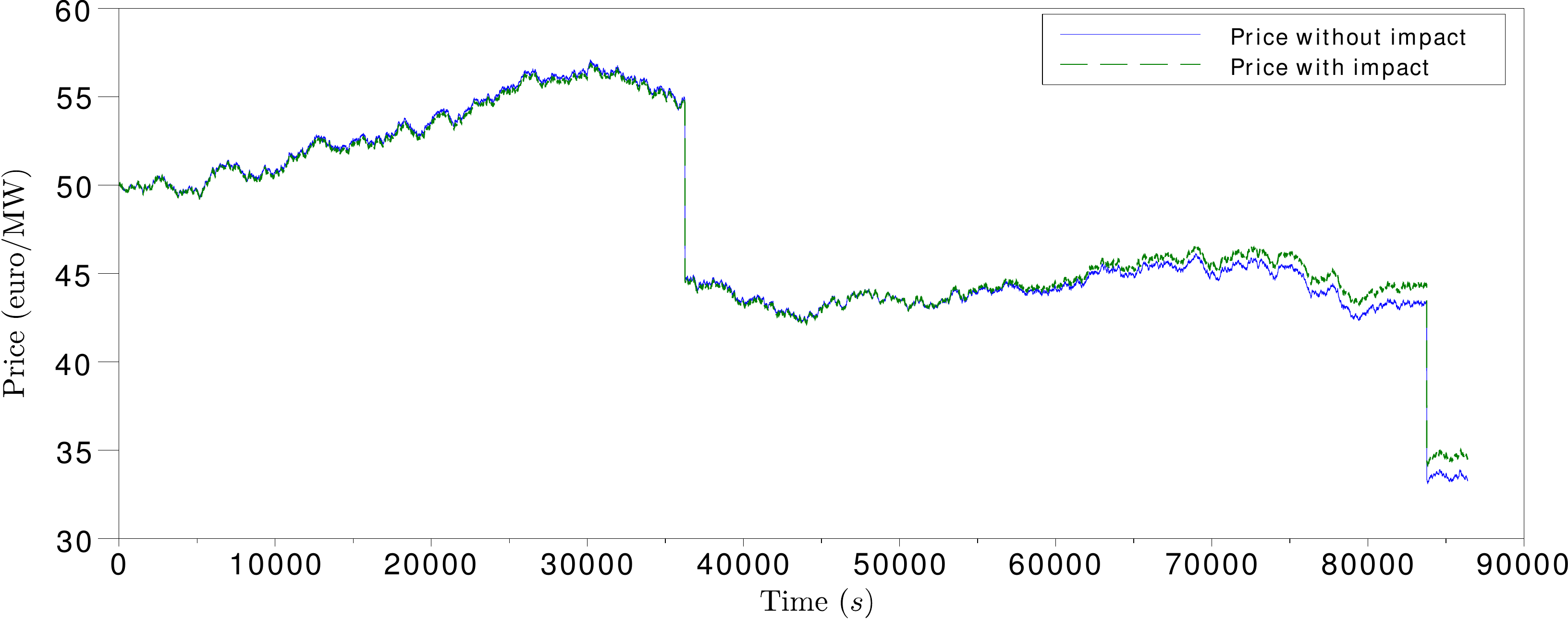}
        \caption{\small{Simulation of the quoted impacted price $\hat Y$ and of the unaffected price $\hat P$}}
        \label{fig:evolutionPrixAvecSansImpactSautsNegatifsDominants}
\end{figure}

\vspace{1cm}

\section{Delay in production}

\setcounter{equation}{0} \setcounter{Assumption}{0}
\setcounter{Theorem}{0} \setcounter{Proposition}{0}
\setcounter{Corollary}{0} \setcounter{Lemma}{0}
\setcounter{Definition}{0} \setcounter{Remark}{0}

In this section, we consider the more realistic situation when there is delay in the production, assumed to be fixed equal to $h$ $\in$ $[0,T]$, and we denote by $v$ $=$ $v_{h}$ the value function to the associated optimal execution problem, as defined in \reff{defvpos}, where we stress  the dependence  in the delay $h$.  Our aim is to show  how one can reduce the problem with delay to a suitable problem without delay, and then solve it explicitly.   We shall consider the problem without jumps on demand forecast and price, but the same argument also works for the case with jumps.

\subsection{Explicit solution with delay}

For simplicity of presentation, and without loss of generality, we  shall focus on  the derivation of the value function $v_h(t,x,y,d)$ for  
an initial time $t$ $=$ $0$,  and fixed  $(x,y,d)$ $\in$ $\R\times\R\times\R$.  Given a control trading rate $q$ $\in$ $\Ac$, and from pathwise uniqueness for the solution to the  dynamics \reff{dynX}, \reff{dynY}, \reff{dynD},  we observe that for any $\xi$ $\in$ $L^0(\Fc_{T-h})$:  
\begin{equation} \label{flowX}
\begin{cases}
X_T^{0,x} + \xi \; = \;  X_T^{T-h,X_{T-h}^{0,x} + \xi} \; a.s.  \\
Y_T^{0,y} \;  = \;  Y_T^{T-h,Y_{T-h}^{0,y}}, \;\;\; D_T^{0,d} \; = \: D_T^{T-h,D_{T-h}^{0,d}} \; a.s. 
\end{cases}
\end{equation}
To alleviate notations, we shall omit the dependence in the fixed initial conditions $(x,y,d)$, and 
simply write $X_s$ $=$ $X_s^{0,x}$, $Y_s$ $=$ $Y_s^{0,y}$,  $D_s$ $=$ $D_s^{0,d}$, for $s$ $\geq$ $0$,  
$v_h$ $=$ $v_h(0,x,y,d)$, and $J(0;q,\xi)$ $=$ $J(0,x,y,d;q,\xi)$ for the the cost functional in \reff{defJ}. 
By the tower property of conditional expectations and from \reff{flowX},
 the cost functional can be written, for all  $q$ $\in$ $\Ac$, $\xi$ $\in$ $L^0(\Fc_{T-h})$, as: 
\beq
& & J(0;q,\xi)  \nonumber \\
&=& \E \Big[ \int_0^{T-h}  q_s \big( Y_s + \gamma q_s) ds + c(\xi) + J(T-h,X_{T-h} + \xi ,Y_{T-h},D_{T-h};q,0) \Big] \label{inegJ} \\
& \geq &   \E \Big[ \int_0^{T-h}  q_s \big( Y_s + \gamma q_s) ds + c(\xi) + v_{_{NP}}(T-h,X_{T-h} + \xi ,Y_{T-h},D_{T-h}) \Big], 
\nonumber 
\enq
by definition \reff{defw0} of the value function $v_{_{NP}}$ for the optimal execution problem without production, i.e. the pure retailer problem.  Since $q$ is arbitrary in $\Ac$, this shows that: 
\beq 
& & \inf_{q \in \Ac} J(0;q,\xi) \label{inegJ2} \\
& \geq &  \inf_{q \in \Ac} \E \Big[ \int_0^{T-h}  q_s \big( Y_s + \gamma q_s) ds + c(\xi) + v_{_{NP}}(T-h,X_{T-h} + \xi ,Y_{T-h},D_{T-h}) \Big], \nonumber 
\enq
for all $\xi$ $\in$ $L^0(\Fc_{T-h})$.  Now, given $q$ $\in$ $\Ac$, and $\xi$ $\in$ $L^0(\Fc_{T-h})$,  let us consider the trading rate 
$\hat q^{NP,\xi}$ in $\Ac_{T-h}$ solution to the pure retailer problem: $v_{_{NP}}(T-h,X_{T-h} + \xi ,Y_{T-h},D_{T-h})$, hence starting at time $T-h$ from 
an inventory $X_{T-h}+\xi$. 
By considering the process $\tilde q$ $\in$ $\Ac$ defined by: $\tilde q_s$ $=$ $q_s$ for $0\leq s< T-h$, and 
$\tilde q_s$ $=$ $\hat q_s^{NP,\xi}$, for $T-h\leq s \leq T$, we then obtain from \reff{inegJ}: 
\beq
& & J(0;\tilde q,\xi) \nonumber \\
&=&  \E \Big[ \int_0^{T-h}  q_s \big( Y_s + \gamma q_s) ds + c(\xi) + v_{_{NP}}(T-h,X_{T-h} + \xi ,Y_{T-h},D_{T-h}) \Big], \label{JNP}
\enq
which proves together with \reff{inegJ2} the equality: 
\beq 
& & \inf_{q \in \Ac} J(0;q,\xi) \label{egJ} \\
& = &  \inf_{q \in \Ac} \E \Big[ \int_0^{T-h}  q_s \big( Y_s + \gamma q_s) ds + c(\xi) + v_{_{NP}}(T-h,X_{T-h} + \xi ,Y_{T-h},D_{T-h}) \Big], \nonumber 
\enq
for all $\xi$ $\in$ $L^0(\Fc_{T-h})$.  Therefore, $v_h$ $=$ $\inf_{q\in\Ac,\xi\in L_+^0(\Fc_{T-h})} J(0;q,\xi)$  can  be written as: 
\beq
v_h  &=&   \inf_{q \in \Ac,\xi \in L_+^0(\Fc_{T-h})} \E \Big[ \int_0^{T-h}  q_s \big( Y_s + \gamma q_s) ds  \nonumber \\
& & \hspace{3cm}  + \;  c(\xi) + v_{_{NP}}(T-h,X_{T-h} + \xi ,Y_{T-h},D_{T-h}) \Big].  \label{valuedelay}
\enq
In other words, the original problem with delay in production is formulated as an optimal execution problem without delay, namely with final horizon 
$T-h$, and terminal cost function: 
\beqs
C_h(x,y,d,\xi) &:=& c(\xi) + v_{_{NP}}(T-h,x+\xi,y,d). 
\enqs
Notice from the explicit expression of $v_{_{NP}}$ in Remark \ref{rembeta2} that this cost function $C_h$ does not depend on $T$, and is in the form:
\beqs
C_h(x,y,d,\xi) &=& C_h(0,y,d-x-\xi,0) \; = \; c(\xi) + v_{_{NP}}(T-h,0,y,d-x-\xi). 
\enqs
The optimization over $q$ and $\xi$ in \reff{valuedelay} is done separately: the production $\xi$ $\in$ $L^0_+(\Fc_{T-h})$ is decided at time $T-h$, after the   choice of the trading rate $(q_s)$ for $0\leq s\leq T-h$ (leading to an inventory $X_{T-h}$), and is determined optimally from the optimization a.s. at $T-h$ of the terminal cost $C_h(X_{T-h},Y_{T-h},D_{T-h},\xi)$. It is then given in feedback form by 
$\xi_{T-h}^*$ $=$  $\hat \xi^{h,+}(D_{T-h} - X_{T-h},Y_{T-h})$ where
\beqs
\hat\xi^{h,+}(d,y) &:=&  \argmin_{\xi \geq 0} C_h(0,y,d-\xi,0) 
\; = \;    \argmin_{\xi \geq 0}  \big[ c(\xi) +  v_{_{NP}}(T-h,0,y,d-\xi) \big],
\enqs
hence explicitly given from the expression of $v_{_{NP}}$ in Remark \ref{rembeta2} by: 
\beq
\hat\xi^{h,+}(d,y) &=& \hat\xi^h(d,y) \ind_{\hat\xi^h(d,y)\geq 0}, \nonumber \\
\hat\xi^h(d,y)  & := &  
\frac{\eta}{\eta+\beta} \Big[ \frac{(\nu h + 2 \gamma)(\mu h + d)  + hy}{(r(\eta,\beta)+\nu)h + 2 \gamma} \Big].  \label{hatxih}
\enq
The problem \reff{valuedelay} is then rewritten as
\beq \label{vhnodelay}
v_h &=& \inf_{q\in\Ac}  \E \Big[ \int_0^{T-h}  q_s \big( Y_s + \gamma q_s) ds  + C_h^+(D_{T-h}-X_{T-h},Y_{T-h}) \Big],
\enq
where
\beqs
C_h^+(d,y) &:=& C_h(0,y,d-\hat\xi^{h,+}(d),0).
\enqs
Notice that when $h$ $=$ $0$, we retrieve the expressions in  the no delay case: 
$\hat\xi^{0,+}$ $=$ $\hat\xi^+$ in \reff{xi+},  $C_0^+$ $=$ $C^+$ in \reff{defC+} and  $v_0$ $=$ $v$ in \reff{defvpos2}.  As in the no delay case, there is no explicit solution to the HJB equation associated to the stochastic control problem \reff{vhnodelay}. We then consider  the approximate control problem where we relax the non negativity constraint on the production, i.e.  $\tilde v_h$ $=$ $\inf_{q\in\Ac,\xi\in L^0(\Fc_{T-h})} J(0;q,\xi)$. Therefore by following the same arguments as above, the corresponding value function is written as:
\beq \label{tildevhnodelay}
\tilde v_h &=& \inf_{q\in\Ac}  \E \Big[ \int_0^{T-h}  q_s \big( Y_s + \gamma q_s) ds  + \tilde C_h(D_{T-h}-X_{T-h},Y_{T-h}) \Big],
\enq
where
\beqs
\tilde C_h(d,y) &:=& C_h(0,y,d-\hat\xi^h(d),0).
\enqs
 From the explicit expressions of $\hat\xi^h$ in \reff{hatxih} and $v_{_{NP}}$ in Remark \ref{rembeta2}, it appears after some tedious but straightforward calculations that the auxiliary terminal cost function $\tilde C_h$ simplifies remarkably  into: 
\beqs
\tilde C_h(d,y) &=&  \tilde v_0(T-h,0,y,d) + K_h,
\enqs
where $\tilde v_0$ is the auxiliary value function without delay explicitly obtained in Theorem~\ref{thm1}, and $K_h$ is a constant depending only on 
the delay $h$ and the parameters of the model, given explicitly by
\beqs
K_h  &= & \frac{\eta^2}{2}\frac{\sigma_0^2+\sigma_d^2\nu^2+2\rho\sigma_0\sigma_d\nu}{(\eta+\beta)(\eta+\nu)(r(\eta,\beta)+\nu)}h\\
        &&\; + \; \gamma\frac{\sigma_0^2+\sigma_d^2\eta^2-2\rho\sigma_0\sigma_d\eta}{(\eta+\nu)^2}\ln\Big(1+\frac{(\eta+\nu)h}{2\gamma}\Big)\\
        && \; - \; \gamma\frac{\sigma_0^2+\sigma_d^2 r^2(\eta,\beta)-2\rho\sigma_0\sigma_dr(\eta,\beta)}{(r(\eta,\beta)+\nu)^2} 
             \ln\Big(1+\frac{(r(\eta,\beta)+\nu)h}{2\gamma}\Big). 
\enqs
One easily checks that $K_h$ $=$ $0$ for $h$ $=$ $0$, and $K_h$ is increasing with $h$ (actually the derivative of $K_h$ w.r.t. $h$ is positive), hence in particular $K_h$ is nonnegative. Plugging into  \reff{tildevhnodelay}, we then get
\beq
\tilde v_h &=& \inf_{q\in\Ac}  \E \Big[ \int_0^{T-h}  q_s \big( Y_s + \gamma q_s) ds + \tilde v_0(T-h,X_{T-h},Y_{T-h},D_{T-h}) \Big]  + K_h.   \label{intertildevh} 
\enq
Therefore, by using the dynamic programming principle for the control problem $\tilde v_0$ $=$ $\tilde v_0(0,x,y,d)$ in \reff{defvaux}, we obtain this 
remarkable  relation 
\beq \label{reldelay}
\tilde v_h &=& \tilde v_0  \; +  \; K_h, 
\enq
which explicitly relates the (approximate) value function with and without delay.  
As expected from the very definition of $\tilde v_h$,  this relation implies that $\tilde v_h-\tilde v_0$ is nonnegative, and is increasing in $h$. This is consistent with the intuition that when making the production choice in advance, we do not take into account the future movements of the price and of the residual demand, which should therefore lead to an average positive correction of the cost.  More precisely, the relation \reff{reldelay} gives  an explicit quantification of the delay impact  via the term $K_h$ (which does not depend on the state variables $x,y,d$)  in function of the various model parameters.  Moreover,  the optimal control  of  the stochastic control problem  \reff{intertildevh} over $[0,T-h)$ is explicitly given by the optimal control $(\hat q_s)_{0\leq s \leq T-h}$ of problem $\tilde v_0$ without delay in Theorem~\ref{thm1}.

\vspace{3mm}

Let us now consider the following strategy $(\hat q^{h,+},\tilde\xi_{T-h}^{h,*})$ $\in$ $\Ac\times L_+^0(\Fc_{T-h})$ for the original 
problem $v_h$ with delay:
\begin{itemize}
\item Before $T-h$, follow the trading strategy $\hat q^{h,+}_s$ $=$ $\hat q_s$, $s$ $<$ $T-h$, corresponding to the solution of the auxiliary problem without delay as if production choice is made at time $T$, and leading to an inventory $\hat X_{T-h}$, and an impacted price $\hat Y_{T-h}$.
\item At time $T-h$, choose the production quantity:
\beqs
\tilde\xi_{T-h}^{h,*} & := & \hat\xi^{h,+}(D_{T-h} - \hat X_{T-h},\hat Y_{T-h}). 
\enqs
\item Between time $T-h$ and $T$, follow the trading strategy $\hat q^{h,+}_s$ $=$ $\hat q_s^{NP,\tilde\xi_{T-h}^{h,*}}$, $T-h\leq s\leq T$, corresponding to the solution of the problem without production, and starting at $T-h$ from an inventory $\hat X_{T-h}$ $+$ $\tilde\xi_{T-h}^{h,*}$. 
\end{itemize}
In order to estimate the quality of this approximate strategy with respect to the optimal trading problem $v_h$, measured by 
\beqs
\Ec_1^h & := &  J(0;\hat q^{h,+},\tilde\xi_{T-h}^{h,*}) - v_h, 
\enqs
we shall compare it with the following strategy  
$(\hat q^h,\hat\xi_{T-h}^h)$  $\in$ $\Ac\times L^0(\Fc_{T-h})$:
\begin{itemize}
\item Before $T-h$, follow the trading strategy $\hat q^{h}_s$ $=$ $\hat q_s$, $s$ $<$ $T-h$,  
corresponding to the solution of the auxiliary problem without delay as if production choice is made at time $T$, and leading to an inventory $\hat X_{T-h}$, and an impacted price $\hat Y_{T-h}$.
\item At time $T-h$, choose the ``production'' quantity (which can be negative):
\beqs
\hat\xi_{T-h}^h &=& \hat\xi^h(D_{T-h} - \hat X_{T-h},\hat Y_{T-h}). 
\enqs
\item Between time $T-h$ and $T$, follow the trading strategy $\hat q^{h}_s$ $=$ $\hat q_s^{NP,\hat\xi_{T-h}^h}$, $T-h\leq s \leq T$, corresponding to the solution of the problem without production, and starting at $T-h$ from an inventory $\hat X_{T-h}$ $+$ $\hat\xi_{T-h}^h$. 
\end{itemize}
Then, by construction and following  the arguments (see in particular \reff{JNP}, \reff{tildevhnodelay}, \reff{intertildevh})  leading  to the expression \reff{reldelay} of $\tilde v_h$, we  see that  $(\hat q^h,\hat\xi_{T-h}^h)$ is the optimal solution for $\tilde v_h$, i.e. $\tilde v_h$ $=$ 
$J(0;\hat q^h,\hat\xi_{T-h}^h)$.  On the other hand, since $\tilde v_h$ $\leq$ $v_h$ $\leq$ $J(0;\hat q^{h,+},\tilde\xi_{T-h}^{h,*})$, we deduce that
\beqs
\max(v_h-\tilde v_h , \Ec_1^h) & \leq &  \bar\Ec^h \; := \;  J(0;\hat q^{h,+},\tilde\xi_{T-h}^{h,*}) - J(0;\hat q^h,\hat\xi_{T-h}^h). 
\enqs
Now, from the expression \reff{JNP} of $J$, and by same arguments as in the proof of Proposition~\ref{erreurnopanne} (see the derivation of relation 
\reff{expressinterEc}), we have
\beqs
\bar\Ec^h &=& \E \Big[ C_h^+(D_{T-h}-\hat X_{T-h},\hat Y_{T-h}) - \tilde C_h(D_{T-h}-\hat X_{T-h},\hat Y_{T-h}) \Big]   \\
&=& \E\Big[ v_{_{NP}}(T-h, 0, \hat{Y}_{T-h}, D_{T-h}-\hat{X}_{T-h}-\tilde{\xi}_{T-h}^{h,*})+c(\tilde{\xi}_{T-h}^{h,*})\\
        &&\;\;\;\;-\; v_{_{NP}}(T-h, 0, \hat{Y}_{T-h}, D_{T-h}-\hat{X}_{T-h}-\hat{\xi}_{T-h}^h)-c(\hat{\xi}_{T-h}^h) \Big]\\
        &=&  \frac{\eta r(\eta,\beta)}{2\beta} \frac{(r(\eta,\beta)+\nu)h+2\gamma}{(\eta+\nu)h+2\gamma} V_h(T) \psi \Big( \frac{m(T,d-x,y)}{\sqrt{V_h(T)}}\Big)
\enqs
where $m$ and $\psi$ are defined as in~\reff{estimerror}, and
\beqs
        V_h(T) &=&  \int_h^T \frac{\sigma_0^2 s^2 + \sigma_d^2(\nu s + 2 \gamma)^2 + 2 \rho\sigma_0\sigma_ds (\nu s + 2 \gamma)} 
{\big[(r(\eta,\beta) + \nu)s + 2 \gamma\big]^2} ds.
\enqs
We recover when $h$ $=$ $0$ the expression in Proposition~\ref{erreurnopanne} of the error in the no delay case, and notice that $\bar\Ec^h$ decreases when the delay increases: indeed, the error comes from the trading procedure before deciding how much to produce, which is dictated by the auxiliary problem, in which the final ``production'' can be negative. After $T-h$, the followed control is optimal, as there remains no production decision at some further date. The shorter the period before making the production decision is, the weaker the error is.

\vspace{1mm}

Let us finally discuss some properties of the (approximate) optimal trading strategy $\hat q^{h,+}$.  
Recalling from Proposition~\ref{propmartinq} that the optimal trading rate is a martingale  in the no delay case, we see by construction of $(\hat q_s^{h,+})_{0\leq s\leq T}$ 
that it is a martingale on $[0,T-h)$ and a martingale on $[T-h,T]$.  Moreover, for any $s$ $\in$ $[T-h,T]$, and $t$ $\in$ $[0,T-h)$, we have
\beq
\E\big[ \hat q_s^{h,+} | \Fc_t \big] &=& \E\big[ \E \big[ \hat q_s^{NP,\tilde\xi_{T-h}^{h,*}}  | \Fc_{T-h} \big]  | \Fc_t  \big] \; = \;  
\E\big[  \hat q_{T-h}^{NP,\tilde\xi_{T-h}^{h,*}}     | \Fc_t  \big] \nonumber \\
&=& \E \Big[ \frac{\eta(\mu h + D_{T-h} - \hat X_{T-h} -  \tilde\xi_{T-h}^{h,*}) - \hat Y_{T-h}}{(\eta+\nu)h + 2\gamma} \big| \Fc_t  \Big] \nonumber \\
&=& \E \Big[ \frac{\eta(\mu h + D_{T-h} - \hat X_{T-h} -  \hat\xi_{T-h}^h) - \hat Y_{T-h}}{(\eta+\nu)h + 2\gamma} \big| \Fc_t  \Big]  \nonumber \\
& & \;\;\; + \;   \frac{\eta}{(\eta+\nu)h + 2\gamma} \E \Big[ \hat\xi_{T-h}^h - \tilde\xi_{T-h}^{h,*} \big| \Fc_t  \Big] \nonumber \\
&=&  \E \Big[ \frac{r(\eta,\beta)(\mu h + D_{T-h} - \hat X_{T-h}) - \hat Y_{T-h}}{(r(\eta,\beta)+\nu)h + 2\gamma} \big| \Fc_t  \Big] \nonumber \\
& & \;\;\; + \;   \frac{\eta}{(\eta+\nu)h + 2\gamma} \E \Big[ \hat\xi_{T-h}^h - \tilde\xi_{T-h}^{h,*} \big| \Fc_t  \Big] \nonumber  \\
&=& \E \big[ \hat q_{T-h} | \Fc_t \big] +    \frac{\eta}{(\eta+\nu)h + 2\gamma} \E \Big[ \hat\xi_{T-h}^h - \tilde\xi_{T-h}^{h,*} \big| \Fc_t  \Big] \nonumber \\
&=& \hat q_t  +    \frac{\eta}{(\eta+\nu)h + 2\gamma} \E \Big[ \hat\xi_{T-h}^h \ind_{\hat\xi_{T-h}^h<0}  \big| \Fc_t  \Big] \; \leq \; \hat q_t \; = \; \hat q_t^{h,+}.   \label{superqh}
\enq
where we used the tower rule for conditional expectations, the martingale property and the explicit expression of $q^{NP,\tilde\xi_{T-h}^{h,*}}$ in Remark \ref{rembeta2},  the  definition of  $\hat\xi_{T-h}^h$, the martingale property and explicit expression of $\hat q$ in  Theorem~\ref{thm1},  and finally the fact that  $\tilde\xi_{T-h}^{h,*}$ $=$ $\hat\xi_{T-h}^h\ind_{\hat\xi_{T-h}^h\geq 0}$.  This shows in particular the supermartingale property of $\hat q^{h,+}$ over the whole period $[0,T]$. Notice that the same arguments as for the derivation of \reff{superqh} shows the martingale property over the whole period 
$[0,T]$ of the optimal trading strategy $\hat q^h$ associated to the auxiliary problem $\tilde v_h$.   
Moreover, by the martingale property of $\hat q^{h,+}$ on $[0,T-h)$, and relation \reff{superqh}, we see that the (approximate) optimal inventory process $\hat X^{h,+}$ with trading rate $\hat q^{h,+}$  has on average, a growth rate  $\frac{d\E[\hat X^{h,+}_s]}{ds}$, which is piecewise constant,  equal to: 
\beqs
\E[\hat q_s^{h,+}] &=& 
\left\{ \begin{array}{ll}
\hat q_0, & \mbox{ for } \; 0 \leq s < T-h \\
\hat q_0^{(h)} := \hat q_0 +   \frac{\eta}{(\eta+\nu)h + 2\gamma} \E \big[ \hat\xi_{T-h}^h \ind_{\hat\xi_{T-h}^h<0}   \big] \; < \; \hat q_0,  &  \mbox{ for }  \; T-h \leq s \leq T, 
\end{array}
\right.
\enqs
with $\hat q_0$ $=$ $\frac{r(\eta,\beta))(\mu T + d-x)-y}{(r(\eta,\beta)+\nu)T +2\gamma}$, and 
\beqs
 \hat q_0^{(h)}   &=&   \hat q_0 -   \frac{\eta r(\eta,\beta)}{\beta((\eta+\nu)h+2\gamma)}\sqrt{V_h(T)}\tilde{\psi}\Big(\frac{m(T,d-x,y)}{\sqrt{V_h(T)}}\Big),
\enqs
where  
\beqs
\tilde{\psi}(z) &:=& \phi(z) -  z\Phi(-z), \;\;\; z \in \R
\enqs
is a nonnegative function, as  pointed out in \reff{relphi}.

\subsection{Numerical results}

We plot figures showing relevant trajectories with the same parameters as in Section~\ref{simul}. We add a delay $h=4$ hours: the production choice has to be made four hours before the end of the trading period. We have
\beqs
\tilde v_h(0,X_0,Y_0,D_0) &=& 1925460 \text{\euro{},}
\enqs
which is slightly higher than the value $\tilde v_0(0,X_0,Y_0,D_0)$ $=$ $1916700 \text{\euro{}}$ without delay.

\begin{figure}[h]
        \centering
        \includegraphics[height=6cm,width=\textwidth]{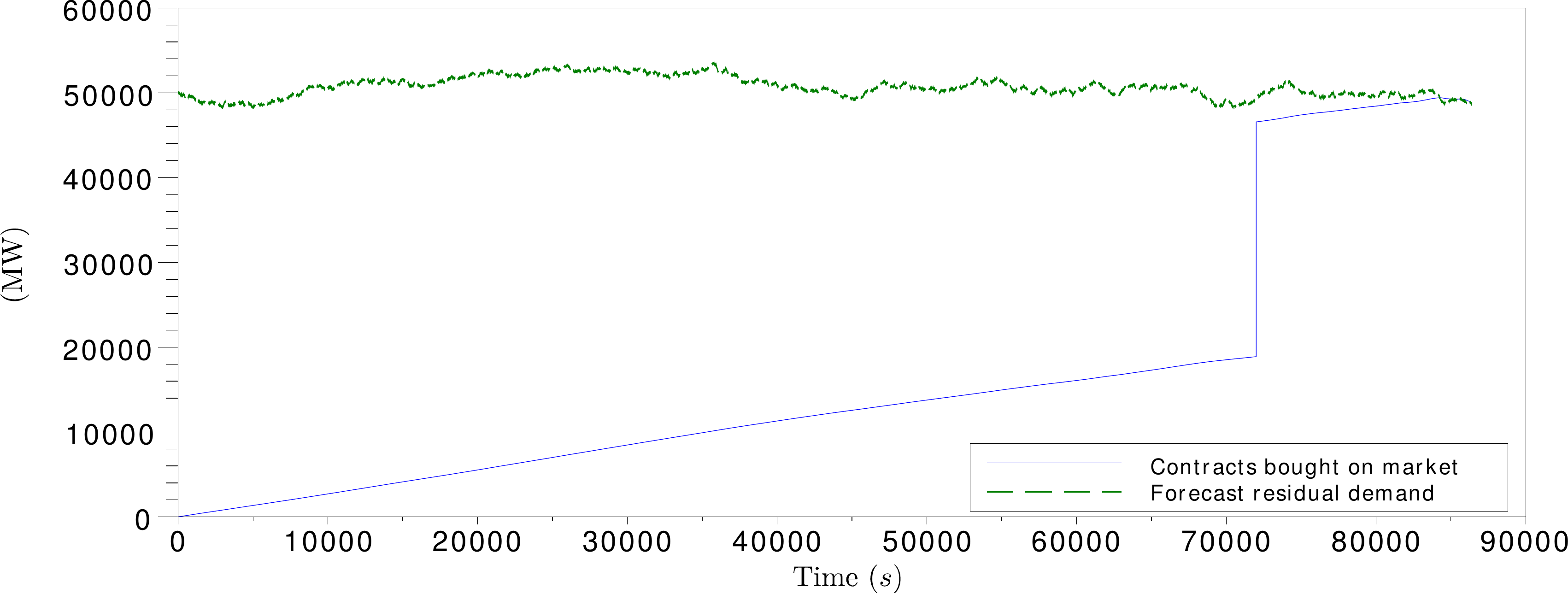}
        \caption{\small{Evolution of the inventory $\hat X$ (with production choice at time $T-h$) and of the forecast residual demand $D$}}
        \label{fig:evolutionXDDelai}
\end{figure}

\begin{figure}[h]
        \centering
        \includegraphics[height=5cm,width=\textwidth]{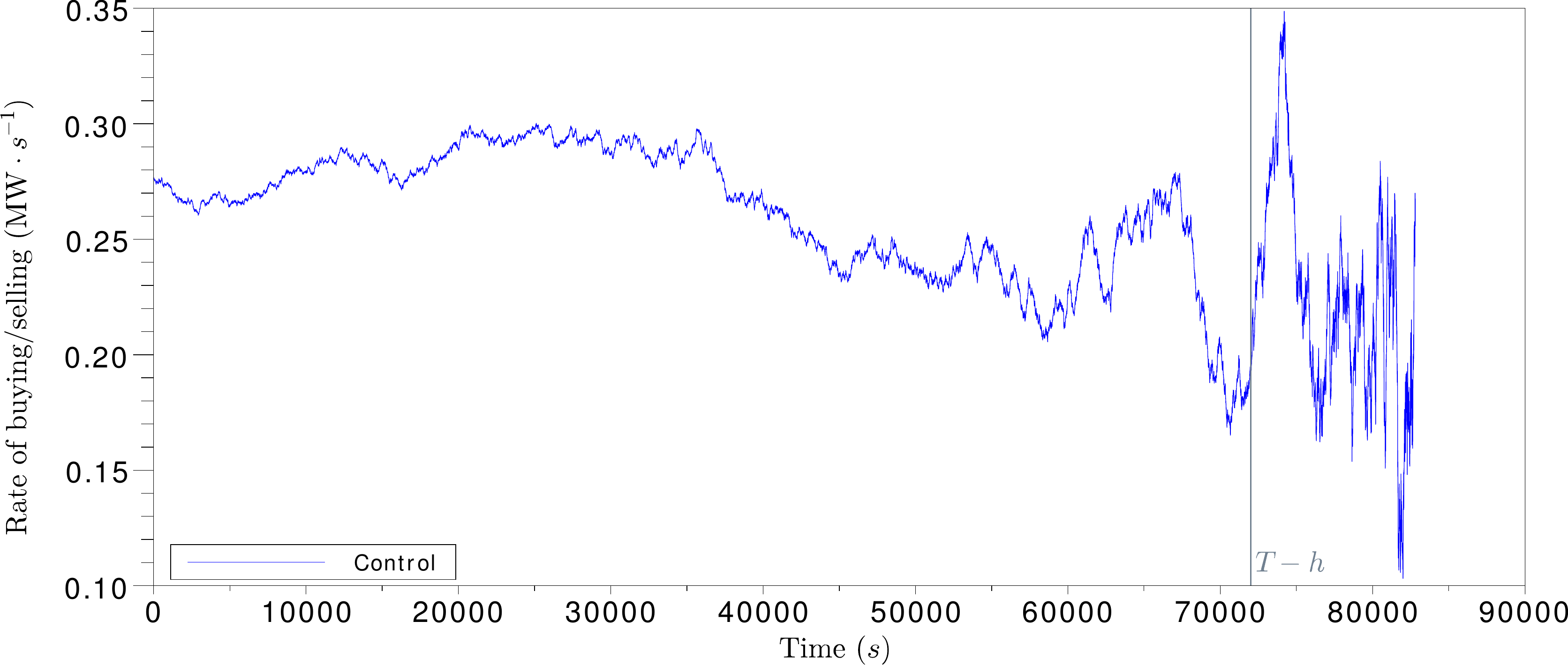}
        \caption{\small{Evolution of the trading rate control $\hat q^{h,+}$ without the last hour}}
        \label{fig:evolutionControleDelaiStopUneHeureAvant}
\end{figure}

On Figure~\ref{fig:evolutionXDDelai}, we see that at time $T-h$, the positive production choice $\tilde{\xi}_{T-h}^{h,*}$ is made, and then we go on buying shares on the intraday market in order to go nearer to the demand forecast, with a smaller slope of trading rate.  
On Figure \ref{fig:evolutionControleDelaiStopUneHeureAvant}, which represents the control process without the last hour of trading (because oscillations then become overwhelming), we see that after date $T-h$, as we do not plan to use final production leverage any more, the approximate optimal control process $\hat q^{h,+}$ oscillates a lot as we are approaching the end of trading time. We can compare with Figure~\ref{fig:evolutionControlePb} to assert that qualitatively, the control in the problem with no production oscillates more than the one in the problem with final production, as in the former problem, the intraday market is the only way to seek to reach the equilibrium.

\vspace{1cm}

\vspace{3mm}

\appendix

\section{Appendix}

\setcounter{equation}{0} \setcounter{Assumption}{0}
\setcounter{Theorem}{0} \setcounter{Proposition}{0}
\setcounter{Corollary}{0} \setcounter{Lemma}{0}
\setcounter{Definition}{0} \setcounter{Remark}{0}

\subsection{Proof of Theorem~\ref{thm1}}

The Hamilton-Jacobi-Bellman (HJB) equation arising from the dynamic programming  associated to the stochastic control problem \reff{defvaux} is: 
\begin{equation}
\begin{cases}
\Dt{\tilde v} + \inf_{q \in \R} \big[ q \Dx{\tilde v} + \nu q \Dy{\tilde v} + \mu \Dd{\tilde v}  +  \frac{1}{2} \sigma_0^2 \Dyy{\tilde v} + \frac{1}{2} \sigma_d^2 \Ddd{\tilde v}  
+ \rho \sigma_0\sigma_d \Dyd{\tilde v}  +   q(y+\gamma q) \big]  \; = \;  0, \nonumber \\
\tilde v(T,x,y,d) \; = \;  \tilde C(d-x) \; = \; \frac{1}{2} r(\eta,\beta) (d-x)^2. \nonumber
\end{cases}
\end{equation}
The argmin in HJB is attained for 
\beqs
\tilde q(t,x,y,d) &=& - \frac{1}{2\gamma}\big[ \Dx{\tilde v} + \nu \Dy{\tilde v} + y \big],  
\enqs
and the HJB equation is rewritten as:
\begin{equation} \label{HJBnojump}
\begin{cases}
\Dt{\tilde v} + \mu \Dd{\tilde v} + \frac{1}{2} \sigma_0^2 \Dyy{\tilde v} + \frac{1}{2} \sigma_d^2 \Ddd{\tilde v}  
+ \rho \sigma_0\sigma_d \Dyd{\tilde v}  - \frac{1}{4\gamma}\big[ \Dx{\tilde v} + \nu \Dy{\tilde v} + y \big]  \; = \;  0,  \\
\tilde v(T,x,y,d)  \; = \; \frac{1}{2} r(\eta,\beta) (d-x)^2.
\end{cases}
\end{equation}
We look for a candidate solution to HJB in the form
\beq
\tilde w(t,x,y,d) &=& A(T-t) (d-x)^2 + B(T-t) y^2 + F(T-t) (d-x)y  \nonumber \\
& & \;\;\; + \;  G(T-t) (d-x) + H(T-t) y + K(T-t), \label{tildewquadra}
\enq
for some deterministic functions $A$, $B$, $F$, $G$, $H$ and $K$.  Plugging the candidate function $\tilde{w}$ into equation \reff{HJBnojump}, we see that $\tilde w$ is solution to the 
HJB equation iff  the following system of ordinary differential  equations (ODEs)  is satisfied by $A$, $B$, $F$, $G$, $H$ and $K$: 
\[
        \left\{ \begin{array}{rll}
        A'+\frac{1}{4\gamma}(-2A+\nu F)^2 &=& 0 \\
        B'+\frac{1}{4\gamma}(2\nu B-F+1)^2 &=& 0 \\
        F'+\frac{1}{2\gamma} (-2A+\nu F) (2\nu B-F+1) &= & 0 \\
        G'-2\mu A +\frac{1}{2\gamma}(-2A+\nu F)(-G+\nu H) &= &0 \\
        H'-\mu F +\frac{1}{2\gamma}(2\nu B-F+1)(-G+\nu H) &=& 0 \\
        K'-\mu G -(\sigma_0^2B+\sigma_d^2 A + \rho\sigma_0\sigma_d F)+\frac{1}{4\gamma}(-G+\nu H)^2 &=& 0
        \end{array} \right.
\]
with the initial conditions $A(0)=\frac{1}{2}r(\eta, \beta)$, $B(0)=0$, $F(0)=0$, $G(0)=0$, $H(0)=0$, $K(0)=0$. We first solve the Riccati system relative to the triple $(A,B,F)$, and obtain: 
\begin{equation}\label{ABF} 
\begin{cases}
        A(t)  \; =  \;  \frac{r(\eta,\beta)(\frac{\nu}{2}t+\gamma)}{(r(\eta,\beta)+\nu)t+2\gamma} \text{,} \\
        B(t) \; = \;  -\frac{1}{2}\frac{t}{(r(\eta,\beta)+\nu)t+2\gamma},  \;\;\;  F(t) \; = \;   \frac{r(\eta,\beta)t}{(r(\eta,\beta)+\nu)t+2\gamma}. 
        \end{cases}
\end{equation}
Then we solve the first-order linear system of ODE relative to the pair  $(G,H)$, which leads to the explicit solution:
\beq \label{GH}
        G(t) \; = \; 2\mu t A(t), \;\;  &\text{ and }& \;  H(t) \; = \; -2 r(\eta,\beta)\mu t B(t). 
\enq
Finally, we explicitly obtain  $K$ from the last equation:
\beq 
        K(t) & =&\gamma\frac{\sigma_0^2 + \sigma_d^2r^2(\eta,\beta) - 2\rho\sigma_0\sigma_dr(\eta,\beta)}{\big(r(\eta,\beta) +  \nu\big)^2} 
			\ln\Big( 1 + \frac{(r(\eta,\beta) +  \nu)t}{2\gamma}\Big)  \nonumber \\
        &  & \;  + \; \frac{\sigma_d^2r(\eta,\beta)\nu+2\rho\sigma_0\sigma_dr(\eta,\beta) - \sigma_0^2}{2\big(r(\eta,\beta) + \nu\big)}t \; 	+ \;  \frac{r(\eta,\beta)\mu^2 t^2(\frac{\nu}{2}t+\gamma)}{ (r(\eta,\beta) + \nu)t+2\gamma}. \label{K}
\enq
By construction,  $\tilde w$ in \reff{tildewquadra} with 
$A$, $B$, $F$, $G$, $H$ and $K$ explicitly given by \reff{ABF}-\reff{GH}-\reff{K}, is a smooth solution with quadratic growth condition to the HJB equation \reff{HJBnojump}. Moreover,  the argmin in HJB equation for $\tilde w$ is attained for
\beqs
\tilde q(t,x,y,d) &=& - \frac{1}{2\gamma}\big[ \Dx{\tilde w} + \nu \Dy{\tilde w} + y \big] \\
&=& \frac{r(\eta,\beta)(\mu(T-t)+d-x)-y}{(r(\eta,\beta)+\nu)(T-t)+2\gamma} \; =: \; \hat q(T-t,d-x,y). 
\enqs
Notice that $\hat q$ is  linear, and  Lipschitz  in $x,y,d$, uniformly in time $t$,  and so given an initial state $(x,y,d)$ at time $t$, 
there exists a unique solution $(\hat X^{t,x,y,d},\hat Y^{t,x,y,d},D^{t,d})_{t\leq s\leq T}$ to \reff{dynX}-\reff{dynY}-\reff{dynD} with the feedback control $\hat q_s$ $=$ $\hat q(T-s,D_s^{t,d}-\hat X_s^{t,x,y,d},\hat Y_s^{t,x,y,d})$, 
which satisfies: $\E[\sup_{t\leq s\leq T} |\hat X_s^{t,x,y,d}|^2 + |\hat Y_s^{t,x,y,d}|^2 + |D_s^{t,d}|^2]$ $<$ $\infty$. This implies in particular that $\E[\int_t^T |\hat q_s|^2 ds]$ $<$ $\infty$, hence $\hat q$ $\in$ $\Ac_t$. 
We now call on a classical verification theorem (see e.g. Theorem~3.5.2 in \cite{bookPham}), which shows that  $\tilde w$ is indeed equal to the value function $\tilde v$, and $\hat q$ is an optimal control. 
Finally, once the optimal trading rate $\hat q$ is determined, the optimal production is obtained from the optimization over $\xi$ $\in$ $\R$  of the terminal cost $C(D_T^{t,d}-\hat X_T^{t,x,y,d},\xi)$, hence given by: 
$\hat\xi_T$ $=$ $\frac{\eta}{\eta+\beta}(D_T^{t,d}-\hat X_T^{t,x,y,d})$. 
\ep

\subsection{Proof of Theorem~\ref{thm2}}

The Hamilton-Jacobi-Bellman (HJB) integro-differential equation arising from the dynamic programming  associated to the stochastic control problem $\tilde v$ $=$ $\tilde v^{(\lambda)}$ with jumps in the dynamics of 
$Y$ and $D$ is: 
\begin{equation}
\begin{cases}
\Dt{\tilde v^{(\lambda)}} + \inf_{q \in \R} \big[ q \Dx{\tilde v^{(\lambda)}} + \nu q \Dy{\tilde v^{(\lambda)}} + \mu \Dd{\tilde v^{(\lambda)}}  +  \frac{1}{2} \sigma_0^2 \Dyy{\tilde v^{(\lambda)}} + \frac{1}{2} \sigma_d^2 \Ddd{\tilde v^{(\lambda)}}  
+ \rho \sigma_0\sigma_d \Dyd{\tilde v^{(\lambda)}}  +   q(y+\gamma q) \big]  \nonumber \\
\; + \;  \lambda \big[ p^+\tilde v^{(\lambda)} (t,x,y+\pi^+,d+\delta^+) +  p^- \tilde v^{(\lambda)}(t,x,y+\pi^-,d+\delta^-) - \tilde v^{(\lambda)}(t,x,y,d) \big] \; = \; 0 \nonumber \\
\tilde v^{(\lambda)}(T,x,y,d) \; = \;  \tilde C(d-x) \; = \; \frac{1}{2} r(\eta,\beta) (d-x)^2. \nonumber
\end{cases}
\end{equation}
Notice that with respect to the no jump case, there is in addition a linear integro-differential term in the HJB equation (which does not depend on the control), and the argmin is attained as in the no jump case for
\beqs
\tilde q^{(\lambda)}(t,x,y,d) &=& - \frac{1}{2\gamma}\big[ \Dx{\tilde v^{(\lambda)}} + \nu \Dy{\tilde v^{(\lambda)}} + y \big]. 
\enqs
The HJB equation is then rewritten as
\begin{equation} \label{HJBjump}
\begin{cases}
\Dt{\tilde v^{(\lambda)}} + \mu \Dd{\tilde v^{(\lambda)}} + \frac{1}{2} \sigma_0^2 \Dyy{\tilde v^{(\lambda)}} + \frac{1}{2} \sigma_d^2 \Ddd{\tilde v^{(\lambda)}}  
+ \rho \sigma_0\sigma_d \Dyd{\tilde v^{(\lambda)}}  - \frac{1}{4\gamma}\big[ \Dx{\tilde v^{(\lambda)}} + \nu \Dy{\tilde v^{(\lambda)}} + y \big]    \\
\; + \;  \lambda \big[ p^+\tilde v^{(\lambda)}(t,x,y+\pi^+,d+\delta^+) +  p^- \tilde v^{(\lambda)}(t,x,y+\pi^-,d+\delta^-) - \tilde v^{(\lambda)}(t,x,y,d) \big] \; = \; 0  \\
\tilde v^{(\lambda)}(T,x,y,d)  \; = \; \frac{1}{2} r(\eta,\beta) (d-x)^2.
\end{cases}
\end{equation}
We look again  for a candidate  solution to \reff{HJBjump} in the form
\beq 
\tilde w^{(\lambda)}(t,x,y,d) &=& A_{\lambda}(T-t) (d-x)^2 + B_{\lambda}(T-t) y^2 + F_{\lambda}(T-t) (d-x)y  \nonumber \\
& & \;\;\; + \;  G_{\lambda}(T-t) (d-x) + H_{\lambda}(T-t) y + K_{\lambda}(T-t), \label{wlambda}
\enq
for some deterministic functions $A_{\lambda}$, $B_{\lambda}$, $F_{\lambda}$, $G_{\lambda}$, $H_{\lambda}$ and $K_{\lambda}$. 
Plugging the candidate function $\tilde w^{(\lambda)}$ into equation \reff{HJBjump}, we see that $\tilde w^{(\lambda)}$ is solution to the 
HJB equation iff  the following system of ordinary differential  equations (ODEs)  is satisfied by $A_{\lambda}$, $B_{\lambda}$, $F_{\lambda}$, $G_{\lambda}$, $H_{\lambda}$ and $K_{\lambda}$:
\[
        \left\{ \begin{array}{rll}
        A_\lambda'+\frac{1}{4\gamma}(-2A_\lambda+\nu F_\lambda)^2 &=& 0 \\
        B_\lambda'+\frac{1}{4\gamma}(2\nu B_\lambda-F_\lambda+1)^2 &=& 0 \\
        F_\lambda'+\frac{1}{2\gamma} (-2A_\lambda+\nu F_\lambda) (2\nu B_\lambda-F_\lambda+1) &= & 0 \\
        G_\lambda'-2\mu A_\lambda +\frac{1}{2\gamma}(-2A_\lambda+\nu F_\lambda)(-G_\lambda+\nu H_\lambda)-\lambda(2\delta A_\lambda+\pi F_\lambda) &= & 0 \\
        H_\lambda'-\mu F_\lambda +\frac{1}{2\gamma}(2\nu B_\lambda-F_\lambda+1)(-G_\lambda+\nu H_\lambda)-\lambda(2\pi B_\lambda +\delta F_\lambda) &= & 0 \\
        K_\lambda'-\mu G_\lambda -(\sigma_0^2 B_\lambda + \sigma_d^2 A_\lambda + \rho\sigma_0\sigma_d F_\lambda)+\frac{1}{4\gamma}(-G_\lambda + \nu H_\lambda)^2& & \\
        \qquad -\lambda[(p^+(\delta^+)^2+p^-(\delta^-)^2) A_\lambda + (p^+(\pi^+)^2+p^-(\pi^-)^2) B_\lambda &&\\
        +(p^+\delta^+\pi^++p^-\delta^-\pi^-) F_\lambda + \delta G_\lambda + \pi H_\lambda] & = &0
        \end{array} \right.
\]
with the initial conditions $A_\lambda(0)=\frac{1}{2}r(\eta, \beta)$, $B_\lambda(0)=0$, $F_\lambda(0)=0$, $G_\lambda(0)=0$, $H_\lambda(0)=0$, $K_\lambda(0)=0$. 
We first solve the Riccati system relative to the triple $(A_\lambda,B_\lambda,F_\lambda)$,  which is  the same as in the no jump case,  and therefore obtain:  
$A_\lambda$ $=$ $A$, $B_\lambda$ $=$ $B$, $F_\lambda$ $=$ $F$ as in \reff{ABF}.  Then we solve the first-order linear system of ODE relative to the pair  $(G_\lambda,H_\lambda)$, which involves the jump parameters 
$\lambda$, $\pi$ and $\delta$, and get:
 \beqs
        G_\lambda(t) &= &  G(t)  +\frac{\lambda}{2}\frac{r(\eta,\beta)t(\pi  t +2\delta  (\nu t+2\gamma))}{(r(\eta,\beta)+\nu)t+2\gamma}, \\ 
        H_\lambda (t) &=&  H(t) -  \frac{\lambda}{2}\frac{(\pi -  2r(\eta,\beta)\delta )t^2}{(r(\eta,\beta)+\nu)t+2\gamma},
\enqs
where $G$ and $H$ are given from  the no jump case \reff{GH}. 
Finally, after some tedious but straightforward calculations,  we explicitly obtain  $K_\lambda$ from the last equation:
\beqs
        K_\lambda(t) &=& K(t)+ \lambda\gamma \frac{p^+(\pi^+ -  r(\eta,\beta)\delta^+)^2+p^-(\pi^- -  r(\eta,\beta)\delta^-)^2} {\big(r(\eta,\beta) + \nu\big)^2} \ln\Big( 1 + \frac{(r(\eta,\beta) +  \nu)t}{2\gamma}\Big) \\
& & - \; \frac{\lambda}{2} \frac{p^+((\pi^+)^2- r(\eta,\beta) \delta^+(2\pi^++\nu\delta^+))+p^-((\pi^-)^2- r(\eta,\beta) \delta^-(2\pi^-+\nu\delta^-))}{r(\eta,\beta) + \nu} t \\
&  &+ \;  \frac{\lambda r(\eta,\beta)}{2}\frac{2\nu\mu\delta  +\lambda((p^+)^2\delta^+(\pi^++\nu\delta^+)+(p^-)^2\delta^-(\pi^-+\nu\delta^-))}{ r(\eta,\beta) + \nu}t^2 \\
&  &+ \; \lambda^2 \gamma r(\eta,\beta) \frac{r(\eta,\beta)\delta^2 +2\nu p^+p^-\delta^+\delta^- -((p^+)^2\delta^+\pi^++(p^-)^2\delta^-\pi^-)}{(r(\eta,\beta)+ \nu)\big((r(\eta,\beta)+\nu)t + 2 \gamma\big)}t^2 \\
&  &+ \; \frac{2\lambda \gamma r(\eta,\beta)^2 \mu\delta}{(r(\eta,\beta)+ \nu)\big((r(\eta,\beta)+\nu)t + 2 \gamma\big)}t^2 - \frac{\lambda^2\pi^2}{48\gamma}t^3 \\
&  &+ \; \frac{\lambda^2p^+p^-r(\eta,\beta)}{2}\frac{2\nu\delta^+\delta^-+\delta^-\pi^++\delta^+\pi^-}{(r(\eta,\beta)+\nu)t + 2\gamma}t^3\\
&  & + \; \frac{1}{8}\frac{4r(\eta,\beta) \mu\lambda\pi  - \lambda^2\pi^2}{(r(\eta,\beta)+\nu)t + 2\gamma}t^3, 
\enqs
with $K$ in \reff{K}. The function $\tilde w^{(\lambda)}$  in \reff{wlambda} may thus be rewritten as the sum of $\tilde w$ in \reff{tildewquadra} and another function of $t$, $d-x$ and $y$, and is by construction a smooth 
solution with quadratic growth condition to the HJB equation \reff{HJBjump}.  Moreover,  the argmin in HJB equation for $\tilde w^{(\lambda)}$ is attained for
\beqs
\tilde q^{(\lambda)}(t,x,y,d) &=& - \frac{1}{2\gamma}\big[ \Dx{\tilde w^{(\lambda)}} + \nu \Dy{\tilde w^{(\lambda)}} + y \big] \\
&=& \frac{r(\eta,\beta)(\mu(T-t)+d-x)-y}{(r(\eta,\beta)+\nu)(T-t)+2\gamma} \\
&  & \;  + \;  \lambda \frac{r(\eta,\beta)\delta  (T-t) + \frac{\pi}{4\gamma}(r(\eta,\beta)+ \nu)(T-t)^2}{(r(\eta,\beta)+\nu)(T-t) + 2\gamma} \\
&  =: &  \hat q^{(\lambda)}(T-t,d-x,y). 
\enqs
Again, notice that $\hat q^{(\lambda)}$ is  linear, and  Lipschitz  in $x,y,d$, uniformly in time $t$,  and so given an initial state $(x,y,d)$ at time $t$, 
there exists a unique solution $(\hat X^{t,x,y,d},\hat Y^{t,x,y,d},D^{t,d})_{t\leq s\leq T}$ to \reff{dynX}-\reff{dynYjump}-\reff{dynDjump} with the feedback control $\hat q_s^{(\lambda)}$ $=$ 
$\hat q^{(\lambda)}(T-s,D_s^{t,d}-\hat X_s^{t,x,y,d},\hat Y_s^{t,x,y,d})$, 
which satisfies: $\E[\sup_{t\leq s\leq T} |\hat X_s^{t,x,y,d}|^2 + |\hat Y_s^{t,x,y,d}|^2 + |D_s^{t,d}|^2]$ $<$ $\infty$, see e.g. Theorem~1.19 in \cite{bookOksendalSulem}. 
This implies  that $\E[\int_t^T |\hat q_s^{(\lambda)}|^2 ds]$ $<$ $\infty$, hence  $\hat q^{(\lambda)}$ $\in$ $\Ac_t$.  We now call on a classical verification theorem for stochastic control of jump-diffusion processes (see e.g. Theorem~3.1 in \cite{bookOksendalSulem}), which shows that   $\tilde w^{(\lambda)}$ is indeed equal to the value function $\tilde v^{(\lambda)}$, and $\hat q^{(\lambda)}$ is an optimal control.  
Finally, once the optimal trading rate $\hat q^{(\lambda)}$ is determined, the optimal production is obtained from the optimization over $\xi$ $\in$ $\R$  of the terminal cost $C(D_T^{t,d}-\hat X_T^{t,x,y,d},\xi)$, hence given by: 
$\hat\xi_T^{(\lambda)}$ $=$ $\frac{\eta}{\eta+\beta}(D_T^{t,d}-\hat X_T^{t,x,y,d})$.  
\ep

\vspace{5mm}

\clearpage

\small

\bibliographystyle{plain}

\end{document}